\documentclass[prd,aps,nofootinbib,tightenlines]{revtex4}
\usepackage{placeins}
\usepackage{xspace}
\usepackage{mathrsfs}
\usepackage{amsmath}
\usepackage{amssymb}
\usepackage{epsfig}
\usepackage{graphicx}
\usepackage{booktabs}
\usepackage{multirow}
\usepackage{subfigure}
\usepackage{bm}
\usepackage{times}
\usepackage{braket}
\usepackage{color}
\usepackage{slashed}
\usepackage{hyperref}
\usepackage{threeparttable}
\definecolor{Red}{rgb}{1.,0.,0.}

\definecolor{Blue}{rgb}{0.,0.,1.}

\definecolor{nicered}{rgb}{0.7,0.1,0.1}
\definecolor{nicegreen}{rgb}{0.1,0.5,0.1}
\hypersetup{colorlinks,citecolor=nicegreen,linkcolor=nicered}
\newcommand{\optbar}[1]{\shortstack{{\tiny (\rule[.4ex]{1em}{.1mm})}\\ [-.7ex] $#1$}}
\newcommand{\beq}{\begin{eqnarray}}
\newcommand{\eeq}{\end{eqnarray}}
\newcommand{\non}{\nonumber\\ }

\newcommand{\KorKbar}{\kern 0.18em\optbar{\kern -0.18em K}{}\xspace}

\newcommand{\epsl}{\epsilon \hspace{-1.6truemm}/\,  }
\newcommand{\psl}{ p \hspace{-2truemm}/ }

\begin{document}
\title{$B_c$ meson decays into $S$-wave charmonium  plus  light meson pairs in the perturbative QCD approach
}
\author{Jia-Ying Wang$^1$}
\author{Jing Jiang$^1$}
\author{Yu-Jie Liu$^1$}\email[Corresponding author:]{liuyujie@njau.edu.cn}
\author{Da-Cheng Yan$^2$}\email[Corresponding author:]{yandac@126.com}
\author{Zhou Rui$^3$}\email[Corresponding author:]{jindui1127@126.com}
\author{Zhen-Jun Xiao$^4$}\email[Corresponding author:]{xiaozhenjun@njnu.edu.cn}
\author{Ya Li$^4$}                \email[Corresponding author:]{liyakelly@163.com}

\affiliation{$^1$ Department of Physics, College of Sciences, Nanjing Agricultural University,
Nanjing, Jiangsu 210095, China}
\affiliation{$^2$ School of Mathematics and Physics, Changzhou University, Changzhou, Jiangsu 213164, China}
\affiliation{$^3$ Department of Physics, Yantai University, Yantai 264005, China}
\affiliation{$^4$ Department of Physics and Institute of Theoretical Physics, Nanjing Normal University, Nanjing 210023, Jiangsu, China}

\date{\today}


\begin{abstract}
In this work, we explore the $P$-wave resonance contributions to the quasi-two-body charmonium decays of $B_c\to \Psi (V\to) P_1P_2$ using the perturbative QCD formalism at leading order, where $\Psi$ denotes a $S$-wave charmonium state, such as $\eta_c(1S,2S),J/\psi$, and $\psi(2S)$. Here, $P_1P_2$ represents a collinear $\pi\pi$ ($K\pi$) pair in the final state, which was primarily produced through the vector resonance $\rho(770)$ ($K^*(892)$ ).
The contributions from the excited $\rho^{\prime}(1450)$, $\rho^{\prime\prime}(1700)$ and $\rho^{\prime\prime\prime}(2254)$ resonances are also taken into account
in the $\pi\pi$  invariant-mass spectrum.
With the improved two-meson distribution amplitudes determined from our previous works,
we examined the $CP$-averaged branching ratios and polarization fractions of the considered quasi-two-body decays.
The longitudinal polarization fractions of the  $B_c\to [J/\psi,\psi(2S)](\rho^+(770)\to)\pi^+\pi^0$ and
$B_c\to [J/\psi,\psi(2S)](K^{*+}\to)K^0\pi^+$ decays are found to be as large as $\sim 90\%$,
since the transverse amplitudes from the dominant factorizable emission diagrams are always power suppressed with respect to the longitudinal ones.
The direct $CP$ violations in $B_c\to \Psi (V\to) P_1P_2$ decays are predicted naturally to be zero as they solely receive contributions from tree diagrams.
Several interesting relative ratios among the branching fractions of the concerned processes are investigated.
In particular,
the obtained ratio $R^{\rm PQCD}_{2\pi/\pi}\equiv \mathcal{B}(B^+_c \to J/\psi\rho^+(770)\to J/\psi \pi^+\pi^0)/{\mathcal{B}(B^+_c \to J/\psi\pi^+)}=2.67^{+0.26}_{-0.18}$ is consistent well with the  LHCb measurement $R^{\rm exp}_{2\pi/\pi}=2.80\pm0.25$.
Other similar ratios proposed in this work can be tested by LHCb experiments in the near future.
\end{abstract}

\maketitle


\section{Introduction}
The $B^+_c$ meson, the lightest $\bar b c$ bound state,
was first discovered by the CDF collaboration \cite{CDF:1998ihx,CDF:1998axz} at the Tevatron collider.
The unique property involving two heavy quarks with different flavors allows either of the heavy quarks can decay with the other behaving as a spectator quark,
or both quarks can annihilate into a virtual $W^+$ boson. Therefore, the weak decays of $B_c$ meson contain rich dynamics in the perturbative regimes, besides the nonperturbative nature, which might shed light on possible new physics beyond the standard model (SM) \cite{Li:2018lxi,Huang:2018nnq,Cheung:2020sbq,Elahi:2021jia}.

The Large Hadron Collider (LHC) experiments have measured abundant hadronic decay channels of $B_c$ meson involving $J/\psi$, $\psi(2S)$ in the final state~\cite{PDG:2024cfk,HFLAV:2024ctg},
which are easily reconstructed experimentally in their dimuon decay modes. For the $\eta_c$ modes, they can be reconstructed using $p\bar{p}$ decay mode as more and more data are accumulated in the future.
However, the exact values of individual branching fractions
for these observed decays are currently unavailable due to the limited information of fragmentation fractions of $\bar b \to B_c$ and the complex experimental background of proton-proton collisions at LHC.
In this context, the measurements of the branching fraction ratios between different $B_c$ decays are befitting since the uncertainties are significantly reduced, allowing for precise tests of the factorization approach.
 To date, the ratios of the branching fraction of $B^+_c$ decaying into charmonium plus an odd (even) number of light hadrons to the reference  $B_c \to J/\psi \pi$  decay mode have been well established~\cite{LHCb:2012ag,LHCb:2013hwj,LHCb:2013rud,CMS:2014oqy,LHCb:2016vni,LHCb:2022ioi,LHCb:2024nlg}.

The ratio $\mathcal{R}_{K/\pi}\equiv \mathcal{B}(B^+_c \to J/\psi K^+)/{\mathcal{B}(B^+_c \to J/\psi\pi^+)}=0.079\pm 0.007\pm 0.003$ was measured with $pp$ collision data at center-of-mass energies of 7 TeV and 8 TeV collected by the LHCb experiment, where the first uncertainty is statistical and the second is systematic \cite{LHCb:2016vni}.
Recently, the LHCb reported the first observation of the $B^+_c \to J/\psi\pi^+\pi^0$ decay \cite{LHCb:2024nlg}.
The ratio of its branching fraction relative to the  $B^+_c \to J/\psi\pi^+$ channel was measured to be $\mathcal{R}^{\rm exp}_{2\pi/\pi}\equiv \mathcal{B}(B^+_c \to J/\psi\pi^+\pi^0)/{\mathcal{B}(B^+_c \to J/\psi\pi^+)}=2.80\pm0.15\pm0.11\pm0.16$, where the first uncertainty is statistical, the second systematic and the third related to imprecise knowledge of the branching factions for $B^+ \to J/\psi K^{*+}$ and $B^+ \to J/\psi K^+$ decays.
The $\pi^+\pi^0$ mass spectrum is found to be consistent with the dominance of an intermediate $\rho^+$ contribution.
Theoretical calculations of $
\mathcal{R}_{K/\pi}$ and $\mathcal{R}_{2\pi/\pi}$ have been carried out using approaches that handle the non-factorizable
and non-perturbative contributions in different ways, yielding values in the range from $0.05$ to $0.10$  and from $2.5$ to $5.7$ respectively \cite{Chang:1992pt,
Liu:1997hr,
Colangelo:1999zn,
AbdEl-Hady:1999jux,
Kiselev:2000pp,Kiselev:2002vz,Ebert:2003cn,
Ivanov:2006ni,
Hernandez:2006gt,
Wang:2007sxa,
Likhoded:2009ib,
Qiao:2012hp,
Naimuddin:2012dy,
Kar:2013fna,
Ke:2013yka,
Rui:2014tpa,
Issadykov:2018myx,
Cheng:2021svx,
Zhang:2023ypl,
Liu:2023kxr}.
Since the vector meson $\rho^+$ decays via the strong interaction with a nontrivial width, the interactions between the resonance and final-state pion pair will show their effects on the branching ratios of the quasi-two-body decays and should be evaluated more precisely.

The decays of $B_c$ mesons into charmonium and light hadrons are expected to be well described by the factorization approach.
Three popular factorization approaches, such as the QCD
factorization (QCDF)~\cite{Beneke:1999br,Beneke:2000ry}, the soft-collinear-effective theory (SCET)~\cite{Bauer:2000yr,Bauer:2001yt,Bauer:2002nz}, and the perturbative QCD (PQCD) factorization~\cite{Keum:2000wi,Lu:2000em,Keum:2000ph}, have been proposed for analyzing two-body hadronic $B$ meson decays.
For most $B \to h_1 h_2$ decay channels, the theoretical predictions obtained by using these different
factorization  approaches agree well with each other and also are well consistent with the data within errors.
The QCDF and SCET are based on the collinear factorization theorem, in which $B$ meson transition form factors develop endpoint singularities if they were computed perturbatively.
Recently, a breakthrough has been made that non-factorizable weak annihilation diagrams in two-body $B$ decays  are calculable in QCDF~\cite{Lu:2022kos}.
The PQCD is based on the $k_T$ factorization theorem, in which the endpoint contribution is absorbed into a transverse-momentum-dependent (TMD) distribution amplitude (DA) or resummed into a Sudakov factor.
In Ref.~\cite{Li:2014xda}, the authors pointed out that the operator-level definition of the transverse-momentum-dependent (TMD)
hadronic wave functions are highly nontrivial in order to avoid the potential light-cone divergence and the rapidity singularity.
A well-defined TMD can be found in Ref.~\cite{Li:2012md}.
Meanwhile, the Sudakov factors from the $k_T$ resummation have been included to suppress the long-distance contributions from the large $b$ region in this work.
The more precise joint resummation derived in~\cite{Li:2013xna} can be included in the future.
For the QCD resummation, one can include its effect as going beyond the tree level in PQCD analysis, which will be done in the future by taking into account the results as given in Refs.~\cite{Li:2012md,Li:2013xna}.

\begin{figure}[h!]
\centerline{\epsfxsize=15cm \epsffile{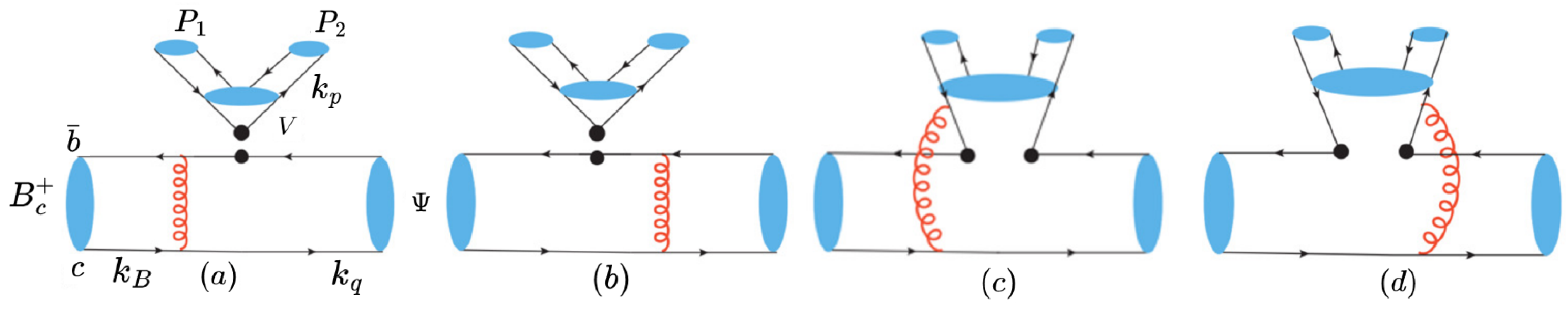}}
\vspace{-0.3cm}
\caption{Typical leading-order Feynman diagrams for the quasi-two-body decays $B_c^+\to \Psi (V\to) P_1P_2$,
 with the charmonium meosn $\Psi=J/\psi, \psi(2S), \eta_c(1S), \eta_c(2S) $, and the
symbol $\bullet$ denotes the weak vertex.}
\label{fey}
\end{figure}

As addressed above, the $B^+_c \to J/\psi\pi^+\pi^0$ decay is expected to
proceed through a $\rho^+ \to \pi^+\pi^0$ intermediate state,  as shown in Fig. \ref{fey}.
In this work, we will study the $P$-wave resonance contributions to quasi-two-body decays $B^+_c \to \Psi(V\to) P_1P_2$ in the PQCD approach,
with the meson pairs $P_1P_2=\pi\pi,K\pi$ and the charmonium $\Psi=\eta_c(1S, 2S), J/\psi, \psi(2S)$.
The invariant mass spectra of the final-state $\pi\pi$ and $K\pi$ pairs are dominated
by the $\rho(770), K^*(892)$, respectively.
The contributions from the excited $\rho^{\prime}(1450)$, $\rho^{\prime\prime}(1700)$ and $\rho^{\prime\prime\prime}(2254)$ resonances are also concerned
in the $\pi\pi$  invariant-mass spectrum.
For simplicity, in the following parts of this work, $\rho, \rho^{\prime}, \rho^{\prime\prime}$ and $\rho^{\prime\prime\prime}$ will be
adopted to represent the  $\rho(770), \rho^{\prime}(1450)$, $\rho^{\prime\prime}(1700)$ and $\rho^{\prime\prime\prime}(2254)$,
respectively.
Currently,
 a factorization formalism that describes multi-body $B$ meson decays in entire phase space is not yet available at present.
It seems reasonable to assume the factorization of three-body $B$ meson decays as a working principle, when two final-state mesons are collimated and the bachelor meson recoils back~\cite{Chen:2002th, El-Bennich:2009gqk, Virto:2016fbw,Krankl:2015fha}.
This situation exists particularly in the low $\pi\pi$ or $K\pi$ invariant mass region ($\lesssim$2 GeV) of the Dalitz plot \cite{Dalitz:1953cp,Dalitz:1954cq} where most resonant structures are seen.
The Dalitz plot is typically dominated by resonant quasi-two-body contributions along the edge.
This proposal provides a theoretical framework for studies of resonant contributions based on the quasi-two-body-decay mechanism.
The nonperturbative (collinear) dynamics responsible for the production of the meson pair, including final-state interactions between the two mesons, is absorbed into two-meson distribution amplitudes (DAs) naturally~\cite{Grozin:1983tt,Grozin:1986at,Muller:1994ses,Diehl:1998dk,Diehl:1998dk1,Diehl:1998dk2}.
The formulation of three-body $B$ meson decays is then simplified to that of quasi-two-body decays, where a Feynman diagram for hard kernels contains a single virtual gluon exchange at leading order (LO) in the strong coupling $\alpha_s$.
The typical PQCD factorization formula for the $B\to P_1P_2P_3$ decay amplitude can be described as the form of \cite{Chen:2002th}
\begin{eqnarray}
\mathcal{A}=\Phi_B\otimes H\otimes \Phi_{P_1P_2}\otimes\Phi_{P_3},
\end{eqnarray}
where the hard kernel $H$ contains only one hard gluon and describes the dynamics of the strong and electroweak interactions in the quasi-two-body hadronic decays as in the formalism for the two-body $B$ meson decays.
The $\Phi_B$ and $\Phi_{P_3}$ are the wave functions for the $B$ meson
and the final state $P_3$, which absorb the non-perturbative dynamics in the relevant  processes.
In recent years, several theoretical approaches have been developed for describing the three-body hadronic decays of $B$ mesons based on the symmetry principles~\cite{Gronau:2013mda,Engelhard:2005hu,Imbeault:2011jz,Xu:2013rua,He:2014xha,Gronau:2005ax}, the QCDF~\cite{Furman:2005xp,El-Bennich:2006rcn,Dedonder:2010fg,Cheng:2007si,Cheng:2013dua,Cheng:2016shb,Li:2014oca,Zhang:2013oqa,Klein:2017xti,Cheng:2020hyj,Qi:2018syl} and the PQCD approaches~\cite{Wang:2016rlo,Li:2016tpn,Ma:2017kec,Li:2018lbd,Rui:2018hls,Wang:2018xux,Xing:2019xti,Li:2018psm,Li:2019pzx,Wang:2020saq,Zou:2020atb,Zou:2020fax,Li:2019hnt,
Li:2020zng,Li:2021cnd}.
Moreover, four-body $B$ meson decays have also been performed in the quasi-two-body framework \cite{Qi:2019nmn,Qi:2020aqz,Rui:2021kbn,Li:2021qiw,Zhang:2021nlw,Yan:2022kck,Yan:2023yvx,Yan:2024ymv,Yan:2025ocu,Liang:2022mrz,Wu:2025dio,Zhang:2025jlt}.

The rest of the paper is organized as follows. The kinematic variables for quasi-two-body hadronic $B$ meson decays are assigned in Sec.~II.
The two-meson $P$-wave DAs are parametrized and normalized to time-like form factors, which take the relativistic Breit-Wigner (RBW) model~\cite{LHCb:2018oeg} or the Gounaris-Sakurai (GS) model~\cite{Gounaris:1968mw}.
We present and discuss the numerical results for branching ratios, polarization fractions and relative ratios of quasi-two-body $B_c$ meson decays in Sec.~III, which is followed by the Conclusion.

\section{THEORETICAL FRAMEWORK}\label{sec:2}

\subsection{Kinematics}

As usual,
we will work in the $B_c$ meson rest frame and choose its momentum $p_{B_c}=\frac{m_{B_c}}{\sqrt{2}}(1,1,\textbf{0}_{\rm T})$ in the light-cone coordinates,
with the ${B_c}$ meson mass $m_{B_c}$.
Considering the quasi-two-body decay $B^+_c \to \Psi (V \to) P_1P_2$,
we define the intermediate vector resonance $V$ with the momentum $p$, and the three final state mesons $\Psi, P_1(P_2)$ with the
momentum $q, p_{1(2)}$, respectively,
obeying the momentum conservation relations $p_{B_c}=p+q$, and $p=p_1+p_2$.
The momentum of  the intermediate resonance $V$ and the final state $\Psi$ can be written as,
\begin{eqnarray}
p=\frac{m_{B_c}}{\sqrt2}(f_{-},f_{+},\textbf{0}_{\rm T}), \quad
q=\frac{m_{B_c}}{\sqrt 2}(g_{+},g_{-},\textbf{0}_{\rm T}),\label{mom-B-k}
\end{eqnarray}
in which the functions $f_{\pm}$ and $g_{\pm}$ read
\begin{eqnarray}
f_{\pm}&=&\frac{1}{2}\left(1+\eta-r_3\pm\sqrt{(1-\eta)^2-2r_3(1+\eta)+r_3^2}\right),\nonumber\\
g_{\pm}&=&\frac{1}{2}\left(1-\eta+r_3\pm\sqrt{(1-\eta)^2-2r_3(1+\eta)+r_3^2}\right),\label{fg}
\end{eqnarray}
with the ratios $r_3=m_{\Psi}^2/m^2_{B_c}$ and $\eta=\omega^2/m^2_{B_c}$,
$m_{\Psi}$ being the mass of charmonium meson $\Psi$  and $\omega^2=p^2$ being the invariant mass squared of the meson pair.
The related longitudinal polarization vectors of the $P$-wave pairs and $J/\psi(\psi(2S))$ meson can then be defined as,
\begin{eqnarray}\label{eq:pq1}
\epsilon_p=\frac{1}{\sqrt{2\eta}}(-f_{-},f_{+},\textbf{0}_{\rm T}),\quad \epsilon_q=\frac{1}{\sqrt{2r_3}}(g_{+},-g_{-},\textbf{0}_{\rm T}),
\end{eqnarray}
satisfying the normalization $\epsilon^2_p=\epsilon^2_q=-1$ and the orthogonality $\epsilon_p \cdot p=\epsilon_q\cdot q=0$.
We also need to define three valence quark momenta labelled by $k_B, k_p, k_q$ in Fig. \ref{fey} to evaluate the hard kernels $H$ in the PQCD approach,
\begin{eqnarray}
 k_{B}=\left(x_Bp_{B_c}^+, x_Bp_{B_c}^- ,\textbf{k}_{B \rm T}\right),
 k_p= \left(zp^+, zp^-,\textbf{k}_{\rm T}\right),
 k_q=\left(x_3q^+, x_3 q^-,\textbf{k}_{3{\rm T}}\right),\label{mom-B-k}
\end{eqnarray}
with the parton momentum fractions (transverse momenta) $x_B, z$ and $x_3$ ($\textbf{k}_{B \rm T}, \textbf{k}_{\rm T}$ and $\textbf{k}_{3{\rm T}}$).

According to the relation $p=p_1+p_2$ and the on-shell conditions $p_{1,2}^{2}=m_{P_{1,2}}^{2}$,
$m_{P_{1,2}}$ being the $P_{1,2}$ meson masses,
we can derive the explicit expressions of the individual momenta $p_{1,2}$,
\begin{eqnarray}\label{eq:p1p2}
 p_1&=&\left((1-\zeta+\frac{r_1-r_2}{2\eta})f_{-}\frac{m_{B_c}}{\sqrt{2}},
 (\zeta+\frac{r_1-r_2}{2\eta})f_{+}\frac{m_{B_c}}{\sqrt{2}}, \textbf{p}_{\rm T}\right), \nonumber\\
 p_2&=&\left((\zeta-\frac{r_1-r_2}{2\eta})f_{-}\frac{m_{B_c}}{\sqrt{2}},
 (1-\zeta-\frac{r_1-r_2}{2\eta})f_{+}\frac{m_{B_c}}{\sqrt{2}}, -\textbf{p}_{\rm T}\right),\nonumber\\
 p_{\rm T}^2&=&\zeta(1-\zeta)\omega^2+\frac{(m_{P_1}^2-m_{P_2}^2)^2}{4\omega^2}-\frac{m^2_{P_1}+m^2_{P_2}}{2},
\end{eqnarray}
with the ratios $r_{1,2}=m_{P_1,P_2}^2/m^2_{B_c}$,
and the factor $\zeta+(r_1-r_2)/(2\eta)=p_1^+/p^+$ characterizing the momentum fraction for one of meson pair up to
corrections from the final state masses.
Alternatively, one can define the polar angle $\theta$ of the meson $P_1$ in the $P_1P_2$ pair rest frame.
The relation between the  momentum fraction $\zeta$ and the polar angle $\theta$ in the dimeson rest frame can be obtained easily,
\begin{eqnarray}\label{eq:cos}
2\zeta-1=\sqrt{1-2\frac{r_1+r_2}{\eta}+\frac{(r_1-r_2)^2}{\eta^2}}\cos\theta,
\end{eqnarray}
with the bounds
\begin{eqnarray}
\zeta_{\text{max,min}}=\frac{1}{2}\left[1\pm\sqrt{1-2\frac{r_1+r_2}{\eta}+\frac{(r_1-r_2)^2}{\eta^2}}\right].
\end{eqnarray}

The relevant branching ratio of the considered quasi-two-body $B_c$ meson decays is given by~\cite{pdg2024}
\begin{eqnarray}\label{eq:br}
\mathcal{B}=\frac{\tau_{B_c} m_{B_c}}{256\pi^3} \int^1_{(\sqrt{r_1}+\sqrt{r_2})^2} d\eta \sqrt{(1-\eta)^2-2r_3(1+\eta)+r^2_3}\int^{\zeta_{\text{max}}}_{\zeta_{\text{min}}}d\zeta|\mathcal{A}|^2,
\end{eqnarray}
with the $B_c$ meson lifetime $\tau_{B_c}$.

Due to the angular momentum conservation requirement, the vector mesons $V$ in the quasi-two-body decays $B_c \to \eta_c(1S,2S)[V\to]P_1P_2$ should be completely polarized in the longitudinal direction.
For $B_c \to [J/\psi,\psi(2S)](V\to)P_1P_2$ decays, both the longitudinal polarization and transverse polarization contribute.
The amplitudes can be decomposed as follows:
\begin{eqnarray}
\mathcal{A}=\mathcal{A}_L+\mathcal{A}_N \epsilon_{T}\cdot \epsilon_{3T}
+i \mathcal{A}_T \epsilon_{\alpha\beta\rho\sigma} n_+^{\alpha} n_-^{\beta} \epsilon_{T}^{\rho} \epsilon_{3T}^{\sigma},
\end{eqnarray}
where $A_L$ is the longitudinally polarized decay amplitude, $A_N$ and $A_T$ are the transversely polarized contributions.
Therefore, the total decay amplitude for $B_c \to [J/\psi,\psi(2S)](V\to)P_1P_2$ decays can be expressed as
\begin{eqnarray}
|\mathcal{A}|^2=|\mathcal{A}_{0}|^2+|\mathcal{A}_{\parallel}|^2+|\mathcal{A}_{\perp}|^2,
\end{eqnarray}
where $\mathcal{A}_0,\mathcal{A}_{\parallel}$ and $\mathcal{A}_{\perp}$ are defined as:
\begin{eqnarray}
\mathcal{A}_0=\mathcal{A}_L, \quad \mathcal{A}_{\parallel}=\sqrt{2}\mathcal{A}_{N},
\quad \mathcal{A}_{\perp}=\sqrt{2}\mathcal{A}_{T}.
\end{eqnarray}
The associated polarization fractions $f_{h}$, $h=0$, $\parallel$ and $\perp$,
are then defined as
\begin{eqnarray}\label{fh}
f_{h}=\frac{{\cal B}_h}{{\cal B}_0+{\cal B}_{||}+{\cal B}_\bot},
\end{eqnarray}
which obey the normalization condition $f_0+f_{\parallel}+f_{\perp}=1$.

\subsection{Distribution Amplitudes}\label{sec:22}
For the wave function of the $B_c$ meson,
we focus on the leading power contribution and adopt its form as being widely used in the PQCD calculations \cite{Keum:2000wi,Liu:2023kxr,Liu:2018kuo,Li:2003yj,Xiao:2011tx},
\begin{eqnarray}
\Phi_{B_c}= \frac{i}{\sqrt{2N_c}} ({ p \hspace{-2.0truemm}/ }_{B_c} +m_{B_c}) \gamma_5 \phi_{B_c} ( x,b), \label{bmeson}
\end{eqnarray}
with the color factor $N_c=3$ and the impact parameter $b$ being conjugate to the parton transverse momentum $k_{B \rm T}$.
The $B_c$ meson DA $\phi_{B_c} (x,b)$ is parametrized as \cite{Liu:2023kxr,Liu:2018kuo}
\beq\label{bcda}
\phi_{B_c}(x,{ b}) &=&   \frac{f_{B_c}}{2\sqrt{2 N_c } }
N_{B_c}x(1-x)\exp\left[-\frac{(1-x)m_c^2+xm_b^2}
{8\beta_{B_c}^2x(1-x)}\right]\exp\left[-2\beta_{B_c}^2x(1-x){ b}^2\right]\;,
\eeq
where $m_b$ and $m_c$ represent  the $b$ quark mass and $c$ quark mass, respectively,
and the shape parameter $\beta_{B_c}$ takes the value $\beta_{B_c}=1.0\pm 0.1$ \cite{Liu:2023kxr,Liu:2018kuo}.
The normalization constant $N_{B_c}$ is related to the $B_c$ meson decay constant  $f_{B_c}$ through the normalization condition
$\int_0^1 \phi_{B_c}(x,b=0)dx=f_{B_c}/(2\sqrt{2N_c})$.

The wave functions of the final-state charmonium meson $\Psi$ can be written as
\beq
\Phi_\Psi(x,b) = \frac{1}{\sqrt{2N_{c}}}\gamma_5
( {p \hspace{-2.0truemm}/ }\phi^{v}(x,b)
+m_{\Psi}\phi^{s}(x,b)),
\eeq
for $\Psi=\eta_c(1S),\eta_c(2S)$ \cite{Li:2017obb,Rui:2015iia},
and
\beq
\Phi_\Psi^{L}(x,b) &=& \frac{1}{\sqrt{2N_{c}}}
\biggl\{m_{\Psi}\epsl_{L}\phi^{L}(x)
+\epsl_{L}\psl\ \phi^{t}(x)\biggl\},
\label{eq:wf-psi-L}\\
\Phi_\Psi^{T}(x,b) &=& \frac{1}{\sqrt{2N_{c}}}
\biggl\{m_{\Psi}\epsl_{T}\phi^{V}(x)
+\epsl_{T}\psl \ \phi^{T}(x)\biggl\},
\label{eq:wf-psi-T}
\eeq
for $\Psi=J/\psi, \psi(2S)$ \cite{Rui:2014tpa,Rui:2015iia}, respectively.
$\epsilon_L$ and $\epsilon_T$  denote the longitudinal and transverse polarization vector of the $J/\psi$ and $\psi(2S)$ mesons.
The explicit forms of the twist-2 DAs $\phi^{v,L,T}$
and the twist-3 DAs $\phi^{s,V,t}$ in the above equations are parameterized as \cite{Li:2017obb,Rui:2014tpa,Rui:2015iia}
\begin{eqnarray}\label{eq:wave}
\phi^{L,T,v}(x,b)&=&\frac{f_{\Psi}}{2\sqrt{2N_c}}N^{L,T,v} x\bar{x}\mathcal {T}(x)
e^{-x\bar{x}\frac{m_c}{\omega_{\Psi}}[\omega_\Psi^2b^2+(\frac{x-\bar{x}}{2x\bar{x}})^2]},\\
\phi^t(x,b)&=&\frac{f_{\Psi}}{2\sqrt{2N_c}}N^t (x-\bar{x})^2\mathcal {T}(x)
e^{-x\bar{x}\frac{m_c}{\omega_\Psi}[\omega_\Psi^2b^2+(\frac{x-\bar{x}}{2x\bar{x}})^2]},\\
\phi^V(x,b)&=&\frac{f_{\Psi}}{2\sqrt{2N_c}}N^V [1+(x-\bar{x})^2]\mathcal {T}(x)
e^{-x\bar{x}\frac{m_c}{\omega_\Psi}[\omega_\Psi^2b^2+(\frac{x-\bar{x}}{2x\bar{x}})^2]},\\
\phi^s(x,b)&=&\frac{f_{\Psi}}{2\sqrt{2N_c}}N^s \mathcal {T}(x)
e^{-x\bar{x}\frac{m_c}{\omega_\Psi}[\omega_\Psi^2b^2+(\frac{x-\bar{x}}{2x\bar{x}})^2]},
\end{eqnarray}
in which  $\mathcal {T}(x)=1$ for $\eta_c(1S), J/\psi$,
and $\mathcal {T}(x)=1-4b^2m_c\omega_\Psi x\bar{x}+\frac{m_c(x-\bar{x})^2}{\omega_\Psi x\bar{x}}$ for $\eta_c(2S), \psi(2S)$ with $\bar{x}=1-x$.
The normalization constant $N^i$ can be determined via the relation $\int_0^1 \phi^i(x,b=0)dx=f_{\Psi}/(2\sqrt{2N_c})$.
The shape parameter $\omega_{\eta_c(1S)}(\omega_{J/\psi})=0.783\pm 0.082$ ($0.667\pm 0.080$) \cite{Dey:2025xdx} is adopted for $\eta_c(1S) (J/\psi)$ meson, and
$\omega_{\eta_c(2S)}(\omega_{\psi(2S)})=0.62\pm 0.02$  for $\eta_c(2S), \psi(2S)$ mesons, which will be discussed in detail in Sec. \ref{NMA}.

The two-meson DA usually depends on the
parton momentum fraction $z$, the meson momentum fraction $\zeta$,
which describes the relative motion between the two mesons in the pair,
and the meson-pair invariant mass squared $\omega^2$.
The $P$-wave two-meson DAs for both longitudinal and transverse polarizations are decomposed, up to twist-3, into~\cite{Wang:2016rlo,Rui:2018hls}
\begin{eqnarray}
\Phi_P^{L}(z,\zeta,\omega)&=&\frac{1}{\sqrt{2N_c}} \left [{ \omega \epsilon\hspace{-1.5truemm}/_p  }\phi_P^0(z,\omega^2)+\omega\phi_P^s(z,\omega^2)
+\frac{{p\hspace{-1.5truemm}/}_1{p\hspace{-1.5truemm}/}_2
  -{p\hspace{-1.5truemm}/}_2{p\hspace{-1.5truemm}/}_1}{\omega(2\zeta-1)}\phi_P^t(z,\omega^2) \right ] (2\zeta-1)\;,\label{pwavel}\\
\Phi_P^{T}(z,\zeta,\omega)&=&\frac{1}{\sqrt{2N_c}}
\Big [\gamma_5{\epsilon\hspace{-1.5truemm}/}_{T}{ p \hspace{-1.5truemm}/ } \phi_P^T(z,\omega^2)
+\omega \gamma_5{\epsilon\hspace{-1.5truemm}/}_{T} \phi_P^a(z,\omega^2)+ i\omega\frac{\epsilon^{\mu\nu\rho\sigma}\gamma_{\mu}
\epsilon_{T\nu}p_{\rho}n_{-\sigma}}{p\cdot n_-} \phi_P^v(z,\omega^2) \Big ]\non
&&\cdot \sqrt{\zeta(1-\zeta)+\alpha}\label{pwavet}\;,
\end{eqnarray}
with the kinematic parameter $\alpha=(r_1-r_2)^2/(4\eta^2)-(r_1+r_2)/(2\eta)$.
The various twists $\phi_P^i$ can be expanded in terms of the Gegenbauer polynomials,
\begin{eqnarray}
\phi_{\pi\pi}^0(z,\omega^2)&=&\frac{3F_{\pi\pi}^{\parallel}(\omega^2)}{\sqrt{2N_c}}z(1-z)\left[1
+a^0_{2\rho}\frac{3}{2}(5t^2-1)\right] \;,\label{eqphi0}\\
\phi_{\pi\pi}^s(z,\omega^2)&=&\frac{3F_{\pi\pi}^{\perp}(\omega^2)}{2\sqrt{2N_c}}t\left[1
+a^s_{2\rho}(10z^2-10z+1)\right]  \;,\label{eqphis} \\
\phi_{\pi\pi}^t(z,\omega^2)&=&\frac{3F_{\pi\pi}^{\perp}(\omega^2)}{2\sqrt{2N_c}}t^2\left[1
+a^t_{2\rho}\frac{3}{2}(5t^2-1)\right]  \;,\label{eqphit}\\
\phi_{\pi\pi}^T(z,\omega^2)&=&\frac{3F_{\pi\pi}^{\perp}(\omega^2)}
{\sqrt{2N_c}}z(1-z)\left[1+a^{T}_{2\rho}\frac{3}{2}(5t^2-1)\right]\;,\label{eqphitt} \\
\phi_{\pi\pi}^a(z,\omega^2)&=&\frac{3F_{\pi\pi}^{\parallel}(\omega^2)}
{4\sqrt{2N_c}}t[1+a_{2\rho}^a(10z^2-10z+1)]\;,\label{eqphia} \\
\phi_{\pi\pi}^v(z,\omega^2)&=&\frac{3F_{\pi\pi}^{\parallel}(\omega^2)}
{8\sqrt{2N_c}}\left[1+t^2+a^v_{2\rho}(3t^2-1)\right]\;,\label{eqphiv}\\
\phi_{K \pi}^0(z,\omega^2)&=&\frac{3F_{K \pi}^{\parallel}(\omega^2)}{\sqrt{2N_c}} z(1-z)\left[1+a_{1K^*}^{||}3t+a_{2K^*}^{||}\frac{3}{2}(5t^2-1)\right]\;,\label{eqphi0kp}\\
\phi_{K \pi}^s(z,\omega^2)&=&\frac{3F_{K \pi}^{\perp}(\omega^2)}{2\sqrt{2N_c}}t\;,\label{eqphiskp}\\
\phi_{K \pi}^t(z,\omega^2)&=&\frac{3F_{K \pi}^{\perp}(\omega^2)}{2\sqrt{2N_c}} t^2\;,\label{eqphitkp}\\
\phi_{K \pi}^T(z,\omega^2)&=&\frac{3F_{K \pi}^{\perp}(\omega^2)}{\sqrt{2N_c}}z(1-z)\left[1+a_{1K^*}^{\perp}3 t+
a_{2K^*}^{\perp}\frac{3}{2}(5 t^2-1)\right]\;,\label{eqphittkp}\\
\phi_{K \pi}^a(z,\omega^2)&=&\frac{3F_{K \pi}^{\parallel}(\omega^2)}{4\sqrt{2N_c}}t\;,\label{eqphiakp}\\
\phi_{K \pi}^v(z,\omega^2)&=&\frac{3F_{K \pi}^{\parallel}(\omega^2)}{8\sqrt{2N_c}}(1+ t^2)\;,\label{eqphivkp}
\end{eqnarray}
where the variable $t=1-2z$ and the Gegenbauer coefficients in the above equations are adopted the same as those determined in Refs. \cite{Yan:2025ocu,Rui:2018hls}:
\begin{eqnarray}
a^0_{2\rho}&=&0.16 \pm 0.10, \quad a^s_{2\rho}=-0.11\pm 0.14,\quad a^t_{2\rho}=-0.21\pm 0.04, \non
a^{T}_{2\rho}&=&0.50\pm 0.50,\quad a^{v}_{2\rho}=-0.50\pm 0.50,\quad a^{a}_{2\rho}=0.40\pm 0.40,\non
a_{1K^*}^{||}&=&0.45\pm 0.11,\quad a_{2K^*}^{||}=-0.75\pm 0.08,\quad a_{1K^*}^{\bot}=0.61\pm 0.21,\quad a_{2K^*}^{\bot}=0.45\pm 0.06.
\label{gegen}
\end{eqnarray}
Note that,
owing to the limited amount of available data, the Gegenbauer coefficients $a^{0,s,t}_{2\rho}$ and $a^{T,a,v}_{2\rho}$ for the excited states $ \rho^{\prime}, \rho^{\prime\prime} \rho^{\prime\prime\prime}$
can not be determined from a global fit to the multi-body $B$ meson decays, as performed for the $\rho(770)$ state in Ref. \cite{Yan:2025ocu}.
Consequently, these coefficients are taken to be the same as those extracted for
the $\rho(770)$ meson in Ref. \cite{Yan:2025ocu}.
For a similar reason,
the twist-3 $K\pi$ DAs $\phi_{K\pi}^{s,t}$ and $\phi_{K\pi}^{a,v}$ have been set to the asymptotic forms in our calculations.
We remark that
the differences among the two-pion DAs corresponding to different intermediate resonances $\rho, \rho^{\prime}, \rho^{\prime\prime},  \rho^{\prime\prime\prime}$
are mainly from the parameters, like masses, widths and the weight coefficients $c_i$, in the time-like form factors,
which will be discussed  in detail in the following paragraphs.

As claimed in Ref.~\cite{Watson:1952ji}, the analysis of
a reaction is expected to be considerably simplified by separating
the effects of the ``primary mechanism" of the reaction and the
``final state interaction". In the present framework, the weak decay of
the $B_c$ meson provides a short-distance production mechanism, while
the strong interaction between the two hadrons in the quasi-two-body
subsystem occurs at a much longer distance scale. According to the
Watson's theorem, the contributions from the short-distance decay
kernel and the long-distance final-state interaction can be separated
effectively in the $B_c$ meson weak decays, and the latter is
absorbed into the two-meson distribution amplitudes. That is, the
aforementioned ``primary mechanism" provided by the nucleon-nucleon
collisions in Ref.~\cite{Watson:1952ji} is replaced by
the weak decays of the $B_c$ meson in the present work. The above
explains why the Watson's theorem can be applied to study the
quasi-two-body $B_c$ meson decays in the PQCD approach.
The rescattering effects in a final-state meson pair can be absorbed into the time-like form factors $F^{\parallel,\perp}(\omega^2)$
according to the Watson theorem~\cite{Watson:1952ji}.
For the narrow intermediate resonance $K^*$,
we employ the RBW line shape to parameterize the form factor $F_{K\pi}^{\parallel}(\omega^2)$
\begin{eqnarray}
\label{BRW}
F^{\parallel}_{K\pi}(\omega^2)&=&\frac{ N_{K\pi}m_{K^*}^2}{m^2_{K^*} -\omega^2-im_{K^*}\Gamma_{K^*}(\omega^2)} \;,
\end{eqnarray}
with $m_{K^*}$ and $\Gamma_{K^*}$ being the pole mass and width of the $K^*$ meson, respectively.
In Ref. \cite{Yan:2025ocu},
an additional coefficient $N_{K\pi}$, associated with the $K\pi$ system corresponding to $K^*(892)$ resonance,
was introduced to remedy the possible theoretical mismatch between the time-like form factors
and the properties of the intermediate $P$-wave resonance.
Moreover, the convergence of the Gegenbauer expansion for the $K\pi$ DAs can be ensured by including the parameter $N_{K\pi}$.
Accordingly, the effect of the parameter $N_{K\pi}=1.48\pm 0.03$
as determined in Ref.~\cite{Yan:2025ocu} is taken into account in the present work as shown in Eq.~(\ref{BRW}).

The mass-dependent width is defined as
\begin{eqnarray}
\label{BRWl}
\Gamma_{K^*}(\omega^2)&=&\Gamma_{K^*}\left(\frac{m_{K^*}}{\omega}\right)\left(\frac{k(\omega)}{k(m_{K^*})}\right)^{(2L_R+1)},
\end{eqnarray}
where the orbital angular momenta $L_R$ in the two-meson system is set to $L_R=1$ for a $P$-wave state.
The $k(\omega)$ is the momentum vector of the decay product measured in the resonance rest frame,
while $k(m_{K^*})$ is the value of $k(\omega)$ at $\omega=m_{K^*}$.
The explicit expression of the $k(\omega)$ is written as
\begin{eqnarray}
k(\omega)=\frac{\sqrt{\lambda(\omega^2,m_{P_1}^2,m_{P_2}^2)}}{2\omega},
\end{eqnarray}
with the K$\ddot{a}$ll$\acute{e}$n function $\lambda (a,b,c)= a^2+b^2+c^2-2(ab+ac+bc)$.

In the experimental analysis of the three-body hadronic $B$ meson decays,
the contribution from a broad $\rho$ resonance is usually parameterized as the GS model~\cite{Gounaris:1968mw}
on basis of the BW function ~\cite{BW-model}.
Taking into account the $\rho$-$\omega$ interference and excited state contributions,
the form factor $F_{\pi\pi}^{||}$ can be written in the form of ~\cite{Li:2016tpn,BaBar:2012bdw,Wang:2016rlo}
\begin{eqnarray}
F^\parallel_{P}(\omega^2)= N_{\pi\pi}\left [ {\rm GS}_\rho(\omega^2,m_{\rho},\Gamma_{\rho})
\frac{1+c_{\omega} {\rm BW}_{\omega}(\omega^2,m_{\omega},\Gamma_{\omega})}{1+c_{\omega}}
+\Sigma c_j {\rm GS}_j(\omega^2,m_j,\Gamma_j)\right] \left( 1+\Sigma c_j\right)^{-1},
\label{GS}
\end{eqnarray}
where $m_{\rho,\omega,j}$ ($\Gamma_{\rho,\omega,j}$), $j=\rho^{\prime}(1450),\rho^{\prime \prime}(1700)$ and $\rho^{\prime \prime \prime}(2254)$, are the masses (decay widths) of the series of $\rho$ resonances, and $c_{\omega,j}$ are the corresponding weights.
Analogously to the $K\pi$ case,
the parameter $N_{\pi\pi}=1.05\pm 0.04$ \cite{Yan:2025ocu}, associated with the $\pi\pi$ system corresponding to $\rho(770)$ resonance, is also introduced.
For the excited $P$-wave resonances $\rho^{\prime}, \rho^{\prime\prime}, \rho^{\prime\prime\prime}$,
we adopt the same value $N_{\pi\pi}=1.05\pm 0.04$ as that associated with the $\rho(770)$ resonance for simplicity.
The masses (GeV) and widths (GeV) for these excited $\rho^{\prime},\rho^{\prime \prime},\rho^{\prime \prime \prime}$ mesons, and the values of the
complex parameters $c_j$ in the above equation can be found in
Ref.~\cite{BaBar:2012bdw}:
\begin{eqnarray}
m_{\rho^\prime}&=&1.493\pm 0.015,\quad\Gamma_{\rho^\prime}=0.427\pm 0.031,\quad c_{\rho^\prime}=(0.158\pm 0.018)e^{i(3.76\pm 0.10)},\label{cr1}\\
m_{\rho^{\prime\prime}}&=&1.861\pm 0.017,\quad \Gamma_{\rho^{\prime\prime}}=0.316\pm 0.026,  \quad c_{\rho^{\prime\prime}}=(0.068\pm 0.009)e^{i(1.39\pm 0.20)},\label{cr2}\\
m_{\rho^{\prime\prime\prime}}&=&2.254\pm 0.022,\quad \Gamma_{\rho^{\prime\prime\prime}}=0.109\pm 0.076,\quad c_{\rho^{\prime\prime\prime}}=(0.0051^{+0.0034}_{-0.0019})e^{i(0.70\pm 0.51)}.\label{cr3}
\end{eqnarray}
The function ${\rm GS}_\rho(\omega^2,m_{\rho},\Gamma_{\rho})$ reads
\begin{equation}
{\rm GS}_\rho(\omega^2, m_\rho, \Gamma_\rho) =
\frac{m_\rho^2 [ 1 + d(m_\rho) \Gamma_\rho/m_\rho ] }{m_\rho^2 - \omega^2 + f(\omega^2, m_\rho, \Gamma_\rho)
- i m_\rho \Gamma (\omega^2, m_\rho, \Gamma_\rho)},
\end{equation}
with the factors
\begin{eqnarray}
\Gamma (s, m_\rho, \Gamma_\rho) &=& \Gamma_\rho  \frac{s}{m_\rho^2}
\left( \frac{\beta_\pi (s) }{ \beta_\pi (m_\rho^2) } \right) ^3~,\non
d(m) &=& \frac{3}{\pi} \frac{m_\pi^2}{g^2(m^2)} \ln \left( \frac{m+2 g(m^2)}{2 m_\pi} \right)
   + \frac{m}{2\pi  g(m^2)}
   - \frac{m_\pi^2  m}{\pi g^3(m^2)}~,\non
f(s, m, \Gamma) &=& \frac{\Gamma  m^2}{g^3(m^2)} \left[ g^2(s) [ h(s)-h(m^2) ]
+ (m^2-s) g^2(m^2)  h'(m^2)\right]~,\non
g(s) &=& \frac{1}{2} \sqrt{s}  \beta_\pi (s)~,\quad
h(s) = \frac{2}{\pi}  \frac{g(s)}{\sqrt{s}}  \ln \left( \frac{\sqrt{s}+2 g(s)}{2 m_\pi} \right),\quad \beta_\pi (s) = \sqrt{1 - 4m_\pi^2/s}.
\end{eqnarray}
Due to the limited investigation on the form factor $F^{\perp}(\omega^2)$,
we usually use the two decay constants $f_V^{(T)}$ of the intermediate vector resonance
to estimate the ratio $F^{\perp}(\omega^2)/F^{\parallel}(\omega^2)\approx f_V^T/f_V$.
\subsection{Decay amplitudes in the PQCD approach}\label{sec:22}
In this section,
we intend to calculate the related decay amplitudes in the PQCD approach.
For the considered quasi-two-body charmonium $B_c^+\to \Psi (V\to)P_1P_2$ decays,
the analytic formulas for the corresponding decay amplitudes can be written in the following form:
\begin{itemize}
\item[(i)] $\Psi=\eta_c(1S), \eta_c(2S)$
\begin{eqnarray}
  {\cal A}(B_c^+\to \eta_c(1S, 2S)[\rho^+\to]\pi^+\pi^0&=&\frac{G_F}{\sqrt{2}}V^*_{cb}V_{ud}\big[(C_2+\frac{C_1}{3})F^{LL}_{e\eta_c}+C_1M^{LL}_{e\eta_c}\big],\label{apetar}\\
  {\cal A}(B_c^+\to \eta_c(1S, 2S)[K^{*+}\to]K^0\pi^+&=&\frac{G_F}{\sqrt{2}}V^*_{cb}V_{us}\big[(C_2+\frac{C_1}{3})F^{LL}_{e\eta_c}+C_1M^{LL}_{e\eta_c}\big],\label{apetak}
\end{eqnarray}
\item[(ii)] $\Psi=J/\psi, \psi(2S)$
\begin{eqnarray}
  {\cal A}^{h}(B_c^+\to [J/\psi, \psi(2S)](\rho^+\to)\pi^+\pi^0&=&\frac{G_F}{\sqrt{2}}V^*_{cb}V_{ud}\big[(C_2+\frac{C_1}{3})F^{LL,h}_{e\psi}+C_1M^{LL,h}_{e\psi}\big],\label{appsir}\\
  {\cal A}^{h}(B_c^+\to [J/\psi, \psi(2S)] (K^{*+}\to)K^0\pi^+&=&\frac{G_F}{\sqrt{2}}V^*_{cb}V_{us}\big[(C_2+\frac{C_1}{3})F^{LL,h}_{e\psi}+C_1M^{LL,h}_{e\psi}\big],\label{appsik}
\end{eqnarray}
\end{itemize}
with the Fermi coupling constant $G_F$,
the CKM matrix elements $V_{cb}, V_{ud(s)}$,
and the Wilson coefficients $C_1$ and $C_2$.
The superscripts $h=0, \|, \bot$ in Eqs.~(\ref{appsir}) and (\ref{appsik}), represent the longitudinal,
parallel, and transverse polarization contributions respectively.
$F_{e\eta_c}^{LL}, F_{e\psi}^{LL, h}$ ($M_{e\eta_c}^{LL}, M_{e\psi}^{LL, h}$)
come from the factorizable (nonfactorizable) emission diagrams as illustrated in Fig. \ref{fey} with the $(V-A)\otimes(V-A)$ current.

The explicit expressions of the individual amplitudes $F_{e\eta_c}^{LL}, F_{e\psi}^{LL, h}$ and
$M_{e\eta_c}^{LL}, M_{e\psi}^{LL, h}$ can be straightforwardly obtained by evaluating the Feyman diagrams in Fig. \ref{fey}.
Performing the standard PQCD calculations,
one gets the following expressions of the relevant amplitudes:
\begin{itemize}
\item[(i)] $B^+_c\to \eta_c(1S,2S)(V\to)P_1P_2$
\begin{eqnarray}
F^{LL}_{e\eta_c}&=& 8\pi C_F m^4_{B_c}F(\omega^2)\int_0^1dx_Bdx_3\int_0^{1/\Lambda}b_Bdb_Bb_3db_3\phi_{B_c}(x_B,b_B)\non
&\times& \{-[(f_-r(-2 + r_b + 2 g_- x_3) - f_+ r (-2 + r_b + 2 g_+ x_3)) \phi^s+(-f_-g_- (-1 + 2 r_b + g_- x3) + f_+g_+ \non
&\times&  (-1 + 2 r_b + g_+ x_3)) \phi^v] E_e(t_a) h_a(\alpha_e,\beta_a,b_B,b_3)S_t(x_3)+[2 r (f_- (g_- + r_c - x_B) -
     f_+ (g_+ + r_c - x_B))\phi^s\non
&+&  (f_+ g_+ (g_- + r_c) -
     f_-g_- (g_+ + r_c) - f_+ g_- x_B + f_- g_+ x_B) \phi^v]h_b(\alpha_e,\beta_a,b_3,b_B)S_t(x_B)\},
\end{eqnarray}
\begin{eqnarray}
M^{LL}_{e\eta_c}&=& \frac{-32\pi C_F m^4_{B_c}}{\sqrt{6}} \int_0^1dx_Bdzdx_3\int_0^{1/\Lambda}b_Bdb_Bbdb\phi_{B_c}(x_B,b_B)\phi_P^0(z)\non
&\times& \{[r (f_-g_- x_3 - f_+g_+ x_3 - f_- x_B + f_+ x_B) \phi^s + (f_- -
     f_+) (-2 g_- g_+ x_3 + (g_- + g_+) x_B + f_-g_- (-1 + z) \non
     &+&
     f_+g_+ (-1 + z)) \phi^v ]E_n(t_c) h_c(\alpha_e,\beta_c,b_B,b)+[r (-f_-g_- x_3 + f_+g_+ x_3 + f_- x_B - f_+ x_B) \phi^s \non
&+&  (f_-g_--
     f_+g_+) ((g_-+ g_+) x_3 - 2 x_B + (f_- + f_+) z) \phi^v]E_n(t_d) h_d(\alpha_e,\beta_d,b_B,b)
\}.
\end{eqnarray}
\item[(ii)] $B^+_c\to [J/\psi, \psi(2S)](V\to)P_1P_2$

The formulas for both the longitudinal and transverse component amplitudes are as follows:
\begin{eqnarray}
F^{LL,0}_{e\psi}&=& 8\pi C_F m^4_{B_c}F(\omega^2)\int_0^1dx_Bdx_3\int_0^{1/\Lambda}b_Bdb_Bb_3db_3\phi_{B_c}(x_B,b_B)\non
&\times& \{-[g_-g_+\phi^v(x_3,b_3)((f_- + f_+)(-2 + r_b)+ 2(f_-g_- + f_+g_+)x_3)-\phi^s(x_3,b_3)r_3(f_-g_-(-1 + 2r_b\non
&+&  g_-x_3) +f_+g_+(-1 + 2r_b + g_+x_3))]  E_e(t_a) h_a(\alpha_e,\beta_a,b_B,b_3)S_t(x_3)+\phi^s(x_3,b_3)[f_+g_+(g_+ + r_c) \non
&-&f_+g_+x_B - f_-g_+x_B]E_e(t_b) h_b(\alpha_e,\beta_b,b_3,b_B)S_t(x_B)\},
\label{f0epsi}
\end{eqnarray}
\begin{eqnarray}
M^{LL,0}_{e\psi}&=& \frac{-32\pi C_F m^4_{B_c}}{\sqrt{6}r_3}\int_0^1dx_Bdzdx_3\int_0^{1/\Lambda}b_Bdb_Bbdb\phi_{B_c}(x_B,b_B)\phi_P^0(z)\non
&\times& \{[g_+g_-\phi^v(x_3,b_B)(f_+(-(g_+x_3) + x_B) +f_-(-g_-x_3 + x_B + 2f_+(-1 + z)))  +(f_- - f_+)\phi^s(x_3,b_B) \non
&\times& r_3(g_-x_B - g_+(x_B + f_+(-1 + z)) + f_-g_-(-1 + z))]E_n(t_c) h_c(\alpha_e,\beta_c,b_B,b)\non
&+& [(f_-g_- + f_+g_+)\phi^s(x_3,b_B)r_3((g_- + g_+)x_3 - 2x_B + (f_- + f_+)z) \non
&+&  g_-g_+\phi^v(x_3,b_B)(f_+(-(g_+x_3) + x_B) + f_-(-(g_-x_3) + x_B - 2f_+z))]E_n(t_d) h_d(\alpha_e,\beta_d,b_B,b)
\},
\end{eqnarray}
\begin{eqnarray}
F^{LL,||}_{e\psi}&=& 8\pi C_F m^4_{B_c}\sqrt{\eta}F(\omega^2)\int_0^1dx_Bdx_3\int_0^{1/\Lambda}b_Bdb_Bb_3db_3\phi_{B_c}(x_B,b_B)\non
&\times& \{-[(-g_+ + g_- (-1 + 4 g_+ x_3)) \phi^T(x_3,b_3) - r (2 + (g_- + g_+) x_3) \phi^V(x_3,b_3)]  E_e(t_a) h_a(\alpha_e,\beta_a,b_B,b_3)S_t(x_3)\non
&+& \phi^V(x_3,b_3)r[(g_- + g_+ + 2 r_c - 2 x_B)]E_e(t_b) h_b(\alpha_e,\beta_b,b_3,b_B)S_t(x_B)\},
\end{eqnarray}
\begin{eqnarray}
F^{LL,\bot}_{e\psi}&=& 8\pi C_F m^4_{B_c}\sqrt{\eta}F(\omega^2)\int_0^1dx_Bdx_3\int_0^{1/\Lambda}b_Bdb_Bb_3db_3\phi_{B_c}(x_B,b_B)\non
&\times& \{-[(g_- - g_+) (\phi^T(x_3,b_3) - r x_3 \phi^V(x_3,b_3))]  E_e(t_a) h_a(\alpha_e,\beta_a,b_B,b_3)S_t(x_3)+ \phi^V(x_3,b_3)r(g_+-g_-)\non
&\times& E_e(t_b) h_b(\alpha_e,\beta_b,b_3,b_B)S_t(x_B)\},
\end{eqnarray}
\begin{eqnarray}
M^{LL,||}_{e\psi}&=& 32\pi C_F m^4_{B_c}\sqrt{\eta}/\sqrt{6}\int_0^1dx_Bdzdx_3\int_0^{1/\Lambda}b_Bdb_Bbdb\phi_{B_c}(x_B,b_B)\non
&\times& \{\phi^T(x_3,b_B)[(-2 g_+g_- x_3 + (g_- + g_+) x_B + f_-g_- (-1 + z) +
    f_+g_+ (-1 + z)) \phi_P^a(z) + ((-g_- + g_+) x_B \non
    &+& f_+g_+ (-1 + z) +
    f_-(g_- - g_- z))\phi^v_P(z)]E_n(t_c) h_c(\alpha_e,\beta_c,b_B,b)\non
&+& [((-2 g_-g_+ x_3 + g_- x_B + g_+ x_B - (f_-g_- + f_+ g_+) z)\phi^T(x_3,b_B) +
    2 r ((g_-+g_+) x_3 -
       2 x_B + (f_-+f_+) z)\non
       &\times& \phi^V(x_3,b_B))\phi_P^a(z)+ (-g_- x_B + g_+ x_B + f_-g_- z -
    f_+g_+ z) \phi^T(x_3,b_B) \phi_P^v(z)]E_n(t_d) h_d(\alpha_e,\beta_d,b_B,b)
\},\non
\end{eqnarray}
\begin{eqnarray}
M^{LL,\bot}_{e\psi}&=& -32\pi C_F m^4_{B_c}\sqrt{\eta}/\sqrt{6}\int_0^1dx_Bdzdx_3\int_0^{1/\Lambda}b_Bdb_Bbdb\phi_{B_c}(x_B,b_B)\non
&\times& \{\phi^T(x_3,b_B)[(g_- x_B - g_+ (x_B + f_+ (-1 + z)) + f_-g_- (-1 + z)) \phi^a_P(x)
 - (-2 g_-g_+ x_3 + (g_- + g_+) x_B \non
 &+& f_-g_- (-1 + z) +
    f_+g_+ (-1 + z)) \phi^a_P(z)]E_n(t_c) h_c(\alpha_e,\beta_c,b_B,b)\non
&+& [(g_- x_B - g_+ x_B - f_- g_- z + f_+g_+ z) \phi^T\phi_P^a
 + ((2 g_+g_- x_3 - g_- x_B - g_+ x_B + f_-g_- z + f_+g_+ z) \phi^T \non
 &-&
    2 r ((g_- + g_+) x_3 -
       2 x_B + (f_-+f_+) z) \phi^V) \phi^v_P]E_n(t_d) h_d(\alpha_e,\beta_d,b_B,b)
\},
\end{eqnarray}
\end{itemize}
where the colour factor $C_F=4/3$, the QCD scale $\Lambda= 0.25 \pm 0.05$ GeV, the mass ratios $r=m_\Psi/m_{B_c}$, $r_{b(c)}=m_{b(c)}/m_{B_c}$,
and the form of the threshold resummation factor $S_t(x)$ can be found in Refs. \cite{Li:2021cnd}.

The evolution factors $E_e(t)$ and $E_n(t)$ in the above factorization formulas are written as
\begin{eqnarray}
E_e(t)&=&\alpha_s(t) \exp[-S_{B_c}(t)-S_\Psi(t)],\nonumber\\
E_n(t)&=&\alpha_s(t) \exp[-S_{B_c}(t)-S_V(t)-S_\Psi(t)],
\end{eqnarray}
where the Sudakov exponents $S_{B_c,V,\Psi}$ are given by  \cite{Liu:2023kxr,Liu:2018kuo}
\begin{eqnarray}
S_{B_c}&=& s(x_B\frac{m_{B_c}}{\sqrt2},b_B )+\frac53\int^t_{m_c}\frac{d\bar\mu}{\bar\mu} \gamma_q(\alpha_s(\bar\mu)),\nonumber\\
S_V&=& s(\frac{m_{B_c}}{\sqrt2} zf_+,b )+ s(\frac{m_{B_c}}{\sqrt2}(1-z)f_+,b )
+ 2\int^t_{1/b}\frac{d\bar\mu}{\bar\mu} \gamma_q(\alpha_s(\bar\mu)),\nonumber\\
S_\Psi&=& s(\frac{m_{B_c}}{\sqrt2}x_3g_+,b_3 ) +s(\frac{m_{B_c}}{\sqrt2}(1-x_3)g_+,b_3 )
+2\int^t_{m_c}\frac{d\bar\mu}{\bar\mu} \gamma_q(\alpha_s(\bar\mu)),
\end{eqnarray}
with the quark anomalous dimension $\gamma_q=-\alpha_s/\pi$.
The explicit expressions of the functions $s(Q,b)$ can be found in the Appendix of Ref.~\cite{Ali:2007ff}.

The hard functions $h_i$, $i=a$-$d$, resulting from the Fourier transformation of virtual quark and
gluon propagators, read
\begin{eqnarray}
h_i(\alpha_e,\beta_i,b_1,b_2)&=&h_1(\alpha_e,b_1) h_2(\beta_i,b_1,b_2),\nonumber\\
h_1(\alpha_e,b_1)&=&\left\{\begin{array}{ll}
K_0(\sqrt{\alpha_e} b_1), & \quad  \quad \alpha_e >0\\
K_0(i\sqrt{-\alpha_e} b_1),& \quad  \quad \alpha_e<0
\end{array} \right.\nonumber\\
h_2(\beta_i,b_1,b_2)&=&\left\{\begin{array}{ll}
\theta(b_1-b_2)I_0(\sqrt{\beta_i}b_2)K_0(\sqrt{\beta_i}b_1)+(b_1\leftrightarrow b_2), & \quad   \beta_i >0\\
\theta(b_1-b_2)I_0(\sqrt{-\beta_i}b_2)K_0(i\sqrt{-\beta_i}b_1)+(b_1\leftrightarrow b_2),& \quad   \beta_i<0
\end{array} \right.
\end{eqnarray}
with the Bessel function $K_0(ix)=\pi[-N_0(x)+iJ_0(x)]$,
and the virtuality $\alpha_e$ ($\beta_i$)  of the internal gluon (quark) in the diagrams:
\begin{eqnarray}
\alpha_e&=&-(g_- x_3 - x_B) (g_+ x_3 - x_B),\non
\beta_a&=&(r_b^2 - (-1 + g_- x_3) (-1 + g_+ x_3)),\non
 \beta_b&=&r_c^2+(g_- - x_B) (-g_+ + x_B)\non
\beta_c&=&-(-g_+ x_3 + x_B + f_- (-1 + z))(-g_- x_3 + x_B + f_+ (-1 + z)),\non
\beta_d&=&-(g_+ x_3 - x_B + f_- z) (g_- x_3 - x_B + f_+ z).
\end{eqnarray}

The hard scales $t_i$, $i=a$-$d$, are chosen as the maximum of the internal
momentum transition in the hard amplitudes:
\begin{eqnarray}
t_{a,b}&=&max\{m_{B_c}\sqrt{|\alpha_e|},m_{B_c}\sqrt{|\beta_{a,b}|},1/b_3,1/b_B\},\non
t_{c,d}&=&max\{m_{B_c}\sqrt{|\alpha_e|},m_{B_c}\sqrt{|\beta_{c,d}|},1/b,1/b_B\}.
\end{eqnarray}
\section{Numerical Analysis} \label{NMA}
In this section,
we first specify the related input parameters in the numerical analysis below,
including the masses, widths (in units of GeV) and  the $B_c$ meson lifetime (in units of ps) \cite{pdg2024},
\begin{eqnarray}
\label{para1}
m_{B_c}&=&6.275,\quad m_{J/\psi}=3.097, \quad m_{\psi(2S)}=3.686,\quad m_{\eta_c(1S)}=2.9834, \quad m_{\eta_c(2S)}=3.6394,\quad m_b=4.8, \non
m_c&=&1.275, \quad~~~~m_{K^\pm}=0.494,\quad
 m_{K^0}=0.498, \quad m_{\pi^0}=0.135,\quad\Gamma_\rho=0.1496, \quad \Gamma_{K^*}=0.0473,\quad
 \tau_{B_c}=0.507,\non
\end{eqnarray}
the decay constants (in units of GeV) \cite{Liu:2023kxr,Liu:2018kuo,Li:2017obb,Rui:2015iia,Li:2016tpn,Ali:2007ff},
\begin{eqnarray}
\label{para2}
f_{B_c}&=&0.489, ~~~\quad f_{J/\psi}=0.405,~~~\quad f_{\psi(2S)}=0.296, ~~~\quad f_{\eta_c(1S)}=0.420, ~~~\quad f_{\eta_c(2S)}=0.243,\non
~~~~~~~~~\quad f_{\rho}&=&0.209 , ~~~~\quad f^T_{\rho}=0.165,
 \quad f_{K^*}=0.217, ~\quad f^T_{K^*}=0.185.
\end{eqnarray}
For the CKM matrix elements,
we adopt the Wolfenstein parametrization with the parameters \cite{pdg2024}: $A=0.790$ and $\lambda=0.2265$.

\begin{table}[htbp!]
	\centering
	\caption{The $CP$ averaged branching ratios of the considered three-body decays $B_c \to \Psi(V\to) P_1P_2$ together with the polarization fractions of $B_c \to [J/\psi,\psi(2S)](V\to) P_1P_2$, with $V=\rho, K^*$, in the LO PQCD approach.
The theoretical uncertainties are attributed to the variations of the shape parameter $\beta_{B_c}$ in the $B_c$ meson DA, of the shape parameter $\omega_\Psi$ in the charmonium meson DA,
of the Gegenbauer moments in the two-meson DAs ,
of the hard scale $t$ and
of the pole mass  and decay width  of the intermediate vector resonance, respectively.
The theoretical uncertainty from the weight coefficients $c_j$ of the $B_c \to \Psi(\rho \to) \pi\pi$ decays
are also concerned in our calculation.
The individual uncertainties are then added in quadrature to get the total errors of the LO PQCD predictions.
}
\begin{ruledtabular}
\begin{threeparttable}
	\begin{tabular}{lccc}
Channels                                              &${\cal B}(10^{-3})$&$f_0(\%)$&$f_\bot(\%)$               \\\hline
$B_c^+ \to J/\psi(\rho^+\to)\pi^+\pi^0$              &$3.15^{+0.90 }_{-0.59}$ &$87.99^{+0.45 }_{-0.09 }$&$1.80\pm 0.08$  \\
$B_c^+ \to J/\psi(K^{*+}\to) K^0 \pi^+$              &$0.12^{+0.03 }_{-0.02 }$&$83.85^{+0.58}_{-0.32 }$& $2.41^{+0.09 }_{-0.13 }$  \\
$B_c^+ \to \psi(2S)(\rho^+\to)\pi^+\pi^0$              &$0.85^{+0.23 }_{-0.19}$ &$84.56^{+0.09 }_{-0.93 }$&$1.83^{+0.14 }_{-0.01 }$  \\
$B_c^+ \to \psi(2S)(K^{*+}\to) K^0 \pi^+$              &$0.030\pm 0.007$  &$77.08^{+0.43 }_{-1.82 }$& $2.64^{+0.24 }_{-0.05}$\\
\hline
$B_c^+ \to \eta_c(1S)(\rho^+\to)\pi^+\pi^0$             &$4.80^{+1.38}_{-0.94}$ &$\cdots$& $\cdots$ \\
$B_c^+ \to \eta_c(1S)(K^{*+}\to) K^0 \pi^+$              &$0.17^{+0.07}_{-0.03 }$  &$\cdots$& $\cdots$   \\
$B_c^+ \to \eta_c(2S)(\rho^+\to)\pi^+\pi^0$              &$0.84^{+0.26}_{-0.18 }$  &$\cdots$& $\cdots$   \\
$B_c^+ \to \eta_c(2S)(K^{*+}\to) K^0 \pi^+$              &$0.028^{+0.009 }_{-0.005 }$  &$\cdots$& $\cdots$  \\
\end{tabular}
\end{threeparttable}
\end{ruledtabular}
\label{brthree}
\end{table}

\begin{table}[htbp!]
	\centering
	\caption{The $CP$ averaged branching ratios of the considered three-body decays $B_c \to \Psi(V\to) P_1P_2$ together with the polarization fractions of $B_c \to [J/\psi,\psi(2S)](V\to) P_1P_2$, with $V=\rho^{\prime}, \rho^{\prime\prime},\rho^{\prime\prime\prime}$, in the LO PQCD approach.
The theoretical uncertainties are attributed to the variations of the shape parameter $\beta_{B_c}$ in the $B_c$ meson DA, of the shape parameter $\omega_\Psi$ in the charmonium meson DA,
of the Gegenbauer moments in the two-meson DAs ,
of the hard scale $t$,
of the pole mass  and decay width  of the intermediate vector resonance
and of the weight coefficients $c_j$,  respectively.
The individual uncertainties are then added in quadrature to get the total errors of the LO PQCD predictions.
}
\begin{ruledtabular}
\begin{threeparttable}
	\begin{tabular}{lccc}
Channels                                              &${\cal B}(10^{-4})$&$f_0(\%)$&$f_\bot(\%)$               \\\hline
$B_c^+ \to J/\psi(\rho^{\prime+}\to)\pi^+\pi^0$              &$4.44^{+1.86}_{-1.46 }$ &$66.99^{+0.65 }_{-0.99 }$&$4.28^{+0.10 }_{-0.23 }$\\
$B_c^+ \to J/\psi(\rho^{\prime\prime+}\to)\pi^+\pi^0$              &$2.60^{+1.10 }_{-0.91 }$ &$59.44^{+1.20 }_{-0.63 }$&$4.85^{+0.26 }_{-0.12 }$\\
$B_c^+ \to J/\psi(\rho^{\prime\prime\prime+}\to)\pi^+\pi^0$              &$0.097^{+0.214 }_{-0.075 }$ &$50.85^{+0.84 }_{-0.80 }$&$5.12^{+0.22 }_{-0.11 }$\\
$B_c^+ \to \psi(2S)(\rho^{\prime+}\to)\pi^+\pi^0$              &$0.88^{+0.35 }_{-0.27 }$ &$62.82^{+0.36 }_{-2.64 }$&$3.60^{+0.17 }_{-0.34 }$ \\
$B_c^+ \to \psi(2S)(\rho^{\prime\prime+}\to)\pi^+\pi^0$              &$0.42^{+0.18 }_{-0.15 }$  &$53.75^{+3.39 }_{-0.00 }$&$3.38^{+0.26 }_{-0.07 }$\\
$B_c^+ \to \psi(2S)(\rho^{\prime\prime\prime+}\to)\pi^+\pi^0$              &$0.0092^{+0.0204}_{-0.0071 }$  &$51.47^{+0.58 }_{-4.55 }$&$2.59^{+0.09 }_{-0.49 }$\\
\hline
$B_c^+ \to \eta_c(1S)(\rho^{\prime+}\to)\pi^+\pi^0$              &$4.40^{+1.85 }_{-1.42 }$ &$\cdots$& $\cdots$\\
$B_c^+ \to \eta_c(1S)(\rho^{\prime\prime+}\to)\pi^+\pi^0$              &$2.03^{+0.86}_{-0.68 }$&$\cdots$& $\cdots$ \\
$B_c^+ \to \eta_c(1S)(\rho^{\prime\prime\prime+}\to)\pi^+\pi^0$              &$0.055^{+0.124 }_{-0.0423 }$&$\cdots$& $\cdots$ \\
$B_c^+ \to \eta_c(2S)(\rho^{\prime+}\to)\pi^+\pi^0$              &$0.52^{+0.22 }_{-0.18 }$ &$\cdots$& $\cdots$\\
$B_c^+ \to \eta_c(2S)(\rho^{\prime\prime+}\to)\pi^+\pi^0$              &$0.18^{+0.07}_{-0.05 }$&$\cdots$& $\cdots$ \\
$B_c^+ \to \eta_c(2S)(\rho^{\prime\prime\prime+}\to)\pi^+\pi^0$              &$0.0020^{+0.0043 }_{-0.0015 }$ &$\cdots$& $\cdots$\\
\end{tabular}
\end{threeparttable}
\end{ruledtabular}
\label{ebrthree}
\end{table}

Based on the decay amplitudes presented in previous section and input parameters above,
we compute the $CP$-averaged branching ratios ($\cal B$) of the quasi-two-body $B_c\to \Psi(V\to)P_1P_2$ decays
in the leading order (LO)  PQCD formalism,
and summarize the predictions in Tables \ref{brthree} and \ref{ebrthree}.
The polarization fractions ($f_0, f_\bot$) of the  $B_c\to [J/\psi,\psi(2S)](V\to) P_1P_2$ decays
are also listed  in Tables \ref{brthree} and \ref{ebrthree}.
It should be noted that, since only tree operators work on these decays,
 the direct $CP$ asymmetries of the considered charmonium quasi-two-body $B_c$ meson decays
are naturally expected to be zero.
If such asymmetries are observed experimentally,
it is probably a signal of new physics.

In our numerical calculations for the branching ratios and polarization fractions,
the first two theoretical uncertainties result from the non-perturbative parameters in the distribution amplitudes of the
initial $B_c$ meson and the final state $\eta_c(1S, 2S)$ and $J/\psi, \psi(2S)$ mesons,
such as the shape parameters $\beta_{B_c}=1.0\pm 0.1$, $\omega_{\eta_c(1S)}(\omega_{J/\psi})=0.783\pm 0.082$ ($0.667\pm 0.080$)  and
$\omega_{\eta_c(2S)}(\omega_{\psi(2S)})=0.62\pm 0.02$.
The third one comes from the Gegenbauer moments in the two-meson $\pi\pi$ and $K\pi$ DAs given in Eqs.~(25)-(36).
The fourth one is caused by the variation of the hard scale $t$ from $0.75t$ to $1.25t$ (without changing $1/b_i$),
which characterizes the effect of the next-to-leading-order QCD contributions.
The fifth and sixth uncertainties are from the pole mass $(m_V)$ and decay width $(\Gamma_V)$  of the intermediate vector resonance.
The last error of the $B_c\to \Psi(V\to)\pi\pi$ decays, with $V=\rho,\rho^{\prime},\rho^{\prime\prime} ,\rho^{\prime\prime\prime}$,
originates from the weight coefficients $c_j$ as shown in Eqs.~(42)-(44).
The individual uncertainties are then added in quadrature to get the total errors of the LO PQCD predictions in Tables \ref{brthree} and \ref{ebrthree}..

Taking five typical decays as examples,
we present the mentioned theoretical uncertainties for the calculated branching ratios $\cal{B}$:
\begin{eqnarray}
{\cal B}(B_c^+ \to J/\psi(\rho^+\to)\pi^+\pi^0)&=(3.15^{+0.83}_{-0.54}(\beta_{B_c})^{+0.30}_{-0.20}(\omega_{J/\psi})^{+0.04}_{-0.01}(a_{2\rho})^{+0.01}_{-0.05}(t)
^{+0.05}_{-0.00}(m_{\rho})^{+0.05}_{-0.00}(\Gamma_{\rho})\nonumber\\
&^{+0.17}_{-0.14}(c_{\rho}))\times 10^{-3},\label{brr}\\
{\cal B}(B_c^+ \to J/\psi(\rho^{\prime+}\to)\pi^+\pi^0)&=(4.44^{+1.23}_{-0.88}(\beta_{B_c})^{+0.43}_{-0.33}(\omega_{J/\psi})^{+0.04}_{-0.05}(a_{2\rho})^{+0.00}_{-0.04}(t)
^{+0.19}_{-0.15}(m_{\rho})^{+0.32}_{-0.31}(\Gamma_{\rho^{\prime}})\nonumber\\
&^{+1.28}_{-1.07}(c_{\rho}))\times 10^{-4},\label{brr1}\\
{\cal B}(B_c^+ \to J/\psi(\rho^{\prime\prime+}\to)\pi^+\pi^0)&=(2.60^{+0.74}_{-0.55}(\beta_{B_c})^{+0.22}_{-0.20}(\omega_{J/\psi})^{+0.00}_{-0.04}(a_{2\rho})^{+0.00}_{-0.05}(t)
^{+0.09}_{-0.12}(m_{\rho})^{+0.22}_{-0.19}(\Gamma_{\rho^{\prime\prime}})\nonumber\\
&^{+0.75}_{-0.65}(c_{\rho}))\times 10^{-4},\\
{\cal B}(B_c^+ \to J/\psi(\rho^{\prime\prime\prime+}\to)\pi^+\pi^0)&=(9.68^{+2.81}_{-2.18}(\beta_{B_c})^{+0.95}_{-0.77}(\omega_{J/\psi})^{+0.09}_{-0.09}(a_{2\rho})^{+0.00}_{-0.19}(t)
^{+0.49}_{-0.48}(m_{\rho})^{+12.52}_{-4.07}(\Gamma_{\rho^{\prime\prime\prime}})\nonumber\\
&^{+16.21}_{-5.56}(c_{\rho}))\times 10^{-6},\label{brr3}\\
{\cal B}(B_c^+ \to J/\psi(K^{*+}\to)K^0\pi^+)&=(0.12^{+0.03}_{-0.02}(\beta_{B_c})^{+0.01}_{-0.01}(\omega_{J/\psi})^{+0.01}_{-0.00}(a_{2K^*})^{+0.00}_{-0.00}(t)
^{+0.01}_{-0.01}(m_{K^*})\nonumber\\
&^{+0.00}_{-0.00}(\Gamma_{K^*}))\times 10^{-3},
\end{eqnarray}
where the third error labeled by $a_{2\rho(2K^*)}$  is derived
by adding in quadrature the individual contributions from the Gegenbauer moments in the two-meson DAs.
The seventh error $c_{\rho}$ is derived
by adding in quadrature the individual contributions from the weight coefficients $c_j$.
In general, the dominant theoretical uncertainties arise from the
shape parameters $\beta_{B_c}$ and $\omega_{J/\psi}$,
which can reach about $30\%$ of the central value of the calculated branching ratio.
For the decay modes involving excited resonances $\rho^{\prime+},\rho^{\prime\prime+},\rho^{\prime\prime\prime+}$,
especially for the channel $B_c^+ \to J/\psi(\rho^{\prime\prime\prime+}\to)\pi^+\pi^0$ shown in Eq.~(\ref{brr3}),
the errors from the variation of the weight coefficient and the decay width
are quite remarkable because the $c_{\rho^{\prime\prime\prime}}=(0.0051^{+0.0034}_{-0.0019})e^{i(0.70\pm 0.51)}$
and $\Gamma_{\rho^{\prime\prime\prime}}=109\pm 79$ (MeV) extracted by the $BABAR$ Collaboration via the $e^+e^-$ annihilation
process~\cite{BaBar:2012bdw} are still poorly constrained.
Therefore, to improve the precision of theoretical predictions for quasi-two-body $B_c$ meson decays, more stringent constraints on the nonperturbative hadronic parameters of the $B_c$ and charmonium mesons, as well as on the parameters associated with the intermediate vector resonances, are required.

The theoretical predictions for three-body hadronic $B_c$ decays in the present work are based on the quasi-two-body approximation,
where the dominant contributions are assumed to arise from intermediate resonant states.
The systematic uncertainty associated with this approximation mainly originates from interference among overlapping resonances,
possible nonresonant contributions, and the limitations of the factorization and final-state-interaction assumptions.
Taking the quasi-two-body decay $B^+_c \to J/\psi\rho^+ \to J/\psi\pi^+\pi^0$ as an example,
we examine the possible systematic uncertainties from the variation of the invariant-mass integration region
around the resonances and the interference among $\rho$ and $\omega$.
We find that the
central values of the ${\cal B}$ are $3.54\times 10^{-3}$ and $3.73\times 10^{-3}$ when the integration
over $\omega$ is limited in the range of $\omega=[m_\rho-1.5\Gamma_\rho,m_\rho+1.5\Gamma_\rho]$
and $\omega=[m_\rho-2\Gamma_\rho,m_\rho+2\Gamma_\rho]$,
and the
resulting uncertainties are at the level of $\mathcal{O}(10\%)$ and $\mathcal{O}(20\%)$, respectively.
The obtained ${\cal B}=3.18\times 10^{-3}$ with the inclusion of the interference among $\rho$ and $\omega$ is actually rather
close to the value $3.15\times 10^{-3}$ in Table \ref{brthree}.
Thus, the systematic uncertainty from the $\rho-\omega$ interference is small,
less than $1\%$.

\begin{figure}[tbp]
\centerline{\epsfxsize=8cm \epsffile{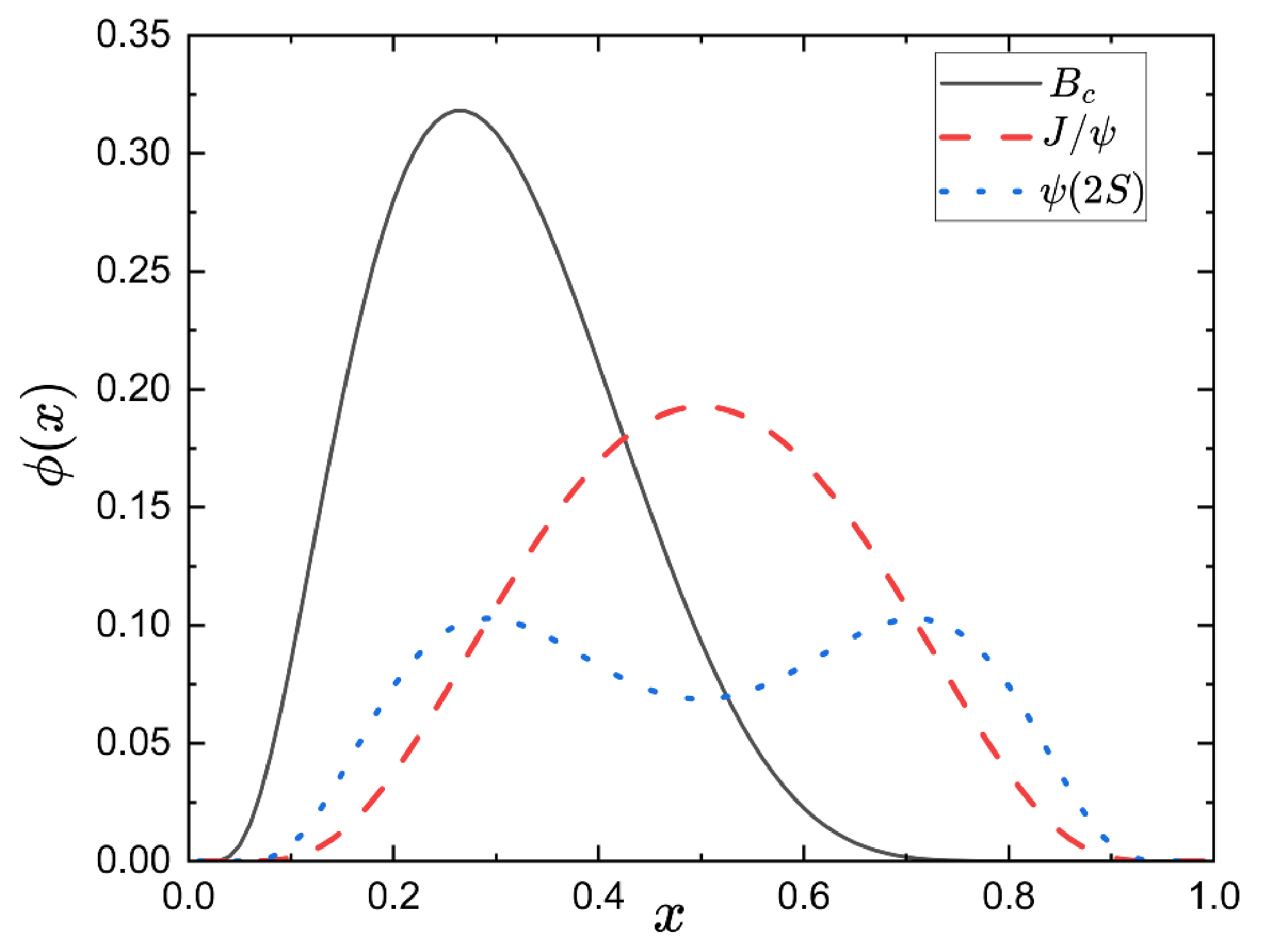}}
\vspace{-0.3cm}
\caption{The overlap of the leading twist distribution amplitudes of the
 initial state $B_c$ (solid line) meson and final state $J/\psi$ (dash line) and $\psi(2S)$ (dot line) mesons at b=0.}
\label{daa}
\end{figure}

From the Table \ref{brthree},
one can observe that the calculated branching fractions of the $B_c^+\to \Psi (K^{*+}\to)K^0\pi^+$ decays
are usually much smaller than those of the $B_c^+\to \Psi (\rho^+\to)\pi^+\pi^0$ ones.
This is predominantly due to the CKM suppression factor $|V_{us}/V_{ud}|^2\sim 0.05$.
In addition,
the ${\cal B}$ of the $B_c$ meson decays to the $2S$ states are found to be smaller
than those of the $1S$ states in the PQCD approach.
This phenomenon can be understood from the wave functions of the $1S$ and $2S$ states.
For example,
we have displayed the shape of the leading twist distribution amplitude of the  $B_c$ meson and the charmonium mesons $J/\psi$ and $\psi(2S)$
in  Fig. \ref{daa}.
It is shown that, relative to the $J/\psi$ state,
the overlap between the initial and final meson wave functions becomes smaller in the $\psi(2S)$ state,
which naturally induces the smaller branching ratio.
Moreover,
the tighter phase space and the smaller decay constants of $2S$ states also result in the suppression of the branching ratios.

The $B_c^+\to [J/\psi, \psi(2S)](V\to)P_1P_2$ decays
are vector-vector modes and can proceed through different polarization amplitudes.
The predicted longitudinal polarization fractions $f_0$ of the four channels  $B_c^+\to [J/\psi, \psi(2S)](\rho^+\to)\pi^+\pi^0$ and $B_c^+\to [J/\psi, \psi(2S)](K^{*+}\to)K^0\pi^+$ can be as large as $\sim 90\%$,
which basically obey the naive counting rules \cite{Li:2004mp}.
In addition,
the calculated $f_0(B_c^+\to J/\psi (\rho^+\to)\pi^+\pi^0)$ and $f_0(B_c^+\to J/\psi (K^{*+}\to)K^0\pi^+)$
also agree well with the previous two-body PQCD analysis in Ref. \cite{Liu:2023kxr}.
For $B_c^+\to [J/\psi, \psi(2S)](V\to)P_1P_2$ decays,
the dominant contributions are from the terms $(C_1/3+C_2)F^{LL, i}_{e\psi}$, $i=0, ||, \bot$,
induced by the tree operators $O_1$ and $O_2$.
Compared with the longitudinal component $F^{LL,0}_{e\psi}$,
the transverse amplitudes $F^{LL,||}_{e\psi}$ and $F^{LL,\bot}_{e\psi}$ of the $B_c^+\to [J/\psi, \psi(2S)](\rho^+\to)\pi^+\pi^0$ and  $B_c^+\to [J/\psi, \psi(2S)](K^{*+}\to)K^0\pi^+$ decays
are highly suppressed by the factors $\sqrt{\eta}=\omega_{\pi\pi}/m_{B_c}\approx m_{\rho}/m_{B_c}\approx 0.12$
and $\sqrt{\eta}=\omega_{K\pi}/m_{B_c}\approx m_{K^*}/m_{B_c}\approx 0.14$, respectively,
which leads  to $f_0\sim 90\%$.
For the excited $\rho^{\prime},\rho^{\prime\prime},\rho^{\prime\prime\prime}$ states,
particularly the $\rho^{\prime\prime\prime}$, however,
the factor $\sqrt{\eta}=\omega_{\pi\pi}/m_{B_c}\approx m_{\rho^{\prime\prime\prime}}/m_{B_c}$ can reach $\sim 0.4$,
which is about three times larger than those of the $\rho$ and $K^*$ resonances.
As a result,
the total transverse polarization fractions of the $B^+_c \to [J/\psi,\psi(2S)](\rho^{\prime\prime\prime+}\to) \pi^+\pi^0$ decays
are almost comparable with the longitudinal ones as shown in Table \ref{ebrthree}.

\begin{figure}[h!]
\centerline{\epsfxsize=15cm \epsffile{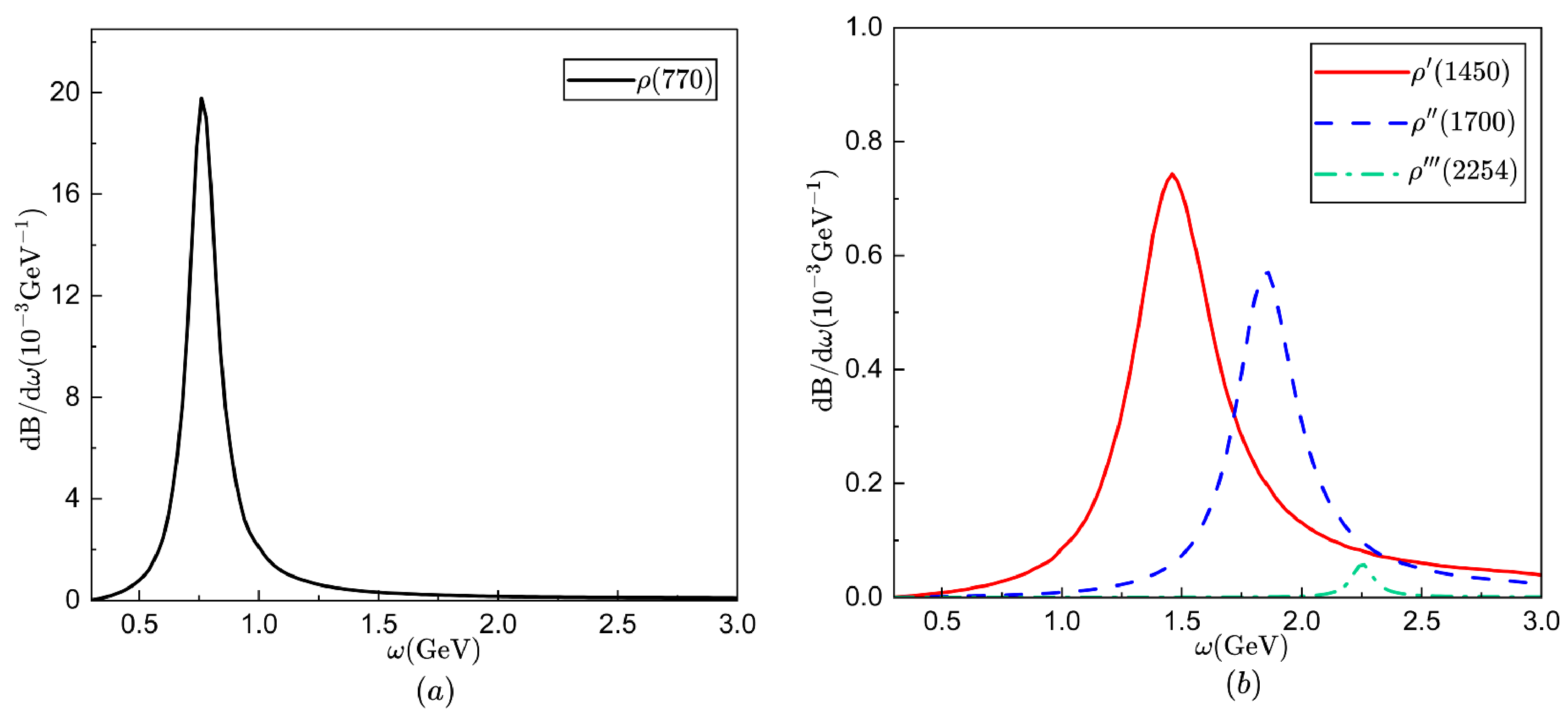}}
\vspace{-0.3cm}
\caption{$(a)$
Differential branching ratio for the $B^+_c\to J/\psi \rho^+\to J/\psi\pi^+\pi^0$ decay,
and
$(b)$
Differential branching ratios for the $B^+_c\to J/\psi \rho^{\prime+}\to J/\psi\pi^+\pi^0$ (solid curve)
,$B^+_c\to J/\psi \rho^{\prime\prime+}\to J/\psi\pi^+\pi^0$ (dashed curve),
and $B^+_c\to J/\psi \rho^{\prime\prime\prime+}\to J/\psi\pi^+\pi^0$ (dotted
curve) decays.}
\label{ws}
\end{figure}

In Fig. \ref{ws}(a),
we present the $\omega$-dependence of the differential decay rate $d{\cal{B}}(B^+_c\to J/\psi \rho^+\to J/\psi\pi^+\pi^0)/d\omega$.
Similarly,
the PQCD prediction of the $d{\cal{B}}/dw$ for the $B^+_c\to J/\psi \rho^{\prime+}\to J/\psi\pi^+\pi^0$ (solid curve)
,$B^+_c\to J/\psi \rho^{\prime\prime+}\to J/\psi\pi^+\pi^0$ (dashed curve),
and $B^+_c\to J/\psi \rho^{\prime\prime\prime+}\to J/\psi\pi^+\pi^0$ (dotted
curve) decays are also displayed in Fig. \ref{ws}(b).
It is shown that the area under the $\rho(770)$ curve in Fig. \ref{ws}(a)
is about one order of magnitude larger than that under the $\rho^{\prime}(1450)$ curve in Fig. \ref{ws}(b),
which basically supports the calculated branching ratios listed in Eqs. (\ref{brr}) and Eq. (\ref{brr1}).
The differential branching ratios exhibit clear peaks at the pole masses of the corresponding resonances.
As expected, the dominant contributions to the branching ratios originate from the regions around the resonances.
The different shapes of the individual channels are mainly governed by the corresponding
GS functions in Eq. (\ref{GS}) and the weight parameters $c_j$ defined in Eqs. (\ref{cr1})-(\ref{cr3}).

\begin{figure}[h!]
\centerline{\epsfxsize=9cm \epsffile{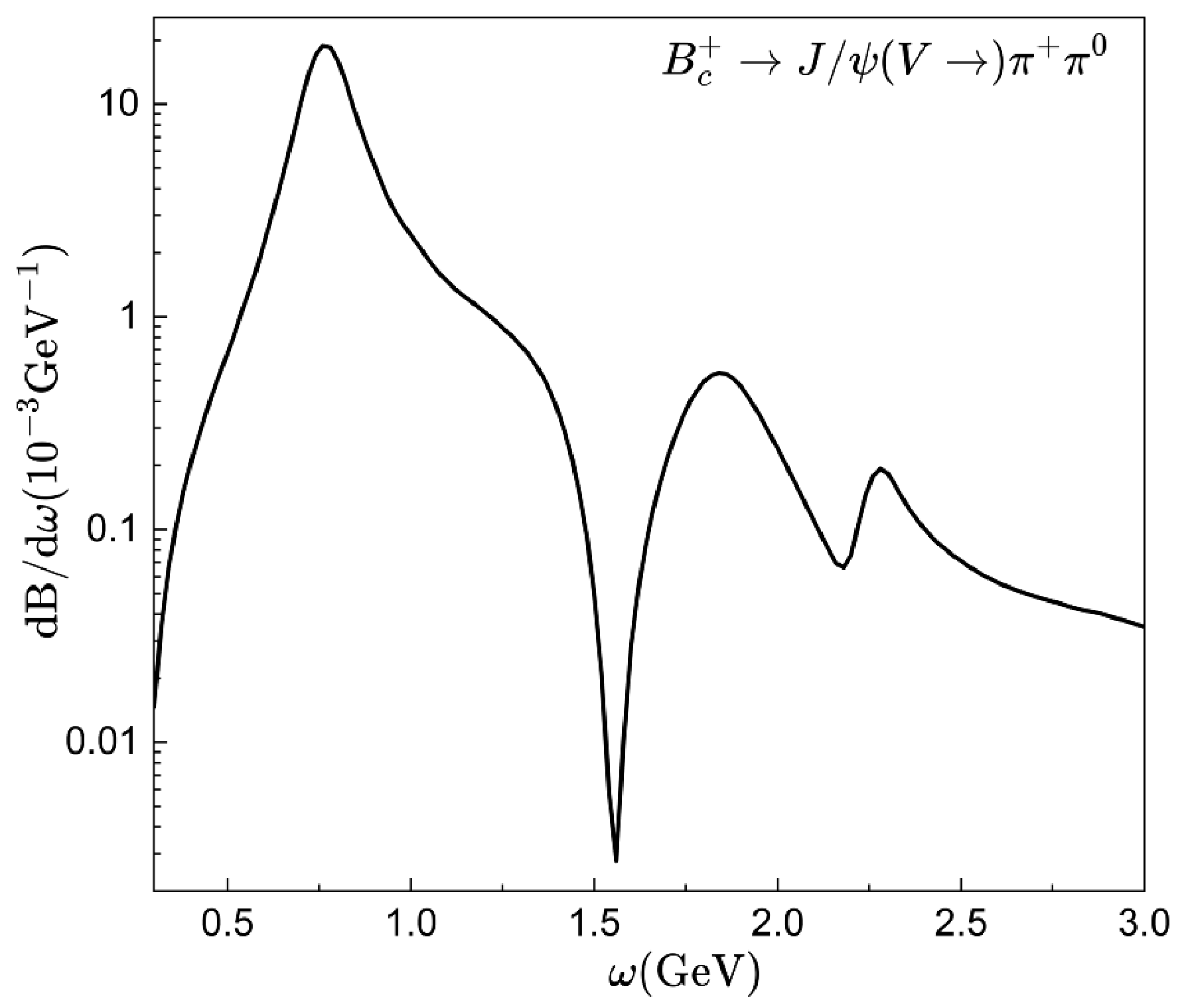}}
\vspace{-0.3cm}
\caption{The differential decay rate of $B_c^+\to J/\psi(V\to) \pi^+\pi^0$ decay with the inclusion of all the contributions
from $\rho(770)$, $\rho^{\prime}(1450)$,
$\rho^{\prime\prime}(1700)$ and $\rho^{\prime\prime\prime}(2254)$ resonances.}
\label{brwa}
\end{figure}

In Fig.~\ref{brwa},
we also plot the total differential branching fraction $d{\cal B}(B_c^+\to J/\psi \pi^+\pi^0)/d\omega$
after the inclusion of the contributions from all the considered $\rho(770)$, $\rho^{\prime}(1450)$,
$\rho^{\prime\prime}(1700)$ and $\rho^{\prime\prime\prime}(2254)$ resonances.
Because the differential decay rate
$d{\cal B}/d\omega$ depends on the values of $|F_\pi(\omega^2)|^2$,
the pattern of the whole curve as shown in Fig.~\ref{brwa}
does agree well with the curve in Fig. 45 of the Ref.~\cite{BaBar:2012bdw},
where the pion form factor-squared $|F_\pi(\omega^2)|^2$ measured by the BABAR Collaboration via the $e^+e^-$ annihilation
process is illustrated as a function of $\sqrt{s^{\prime}}$ (i.e. $m(\pi\pi)$) in the region from 0.3 to 3 GeV.
It is interesting to see a clear dip appearing at the invariant mass around $1.5\sim 1.6$ GeV in Fig.~\ref{brwa},
which is usually interpreted as the strong destructive interference between the $\rho^{\prime}(1450)$ and $\rho^{\prime\prime}(1700)$ channels~\cite{Li:2017mao}.
To be more specific,
we present the PQCD predictions for the individual decay rates ${\cal B}(B_c^+ \to J/\psi(\rho^{\prime+}\to)\pi^+\pi^0)$ and
${\cal B}(B_c^+ \to J/\psi(\rho^{\prime\prime+}\to)\pi^+\pi^0)$,  and the interference term as follows,
\begin{eqnarray}
{\cal B}(B_c^+ \to J/\psi(\rho^{\prime+}\to)\pi^+\pi^0)&\approx& 4.44\times 10^{-4},\nonumber\\ 
{\cal B}(B_c^+ \to J/\psi(\rho^{\prime\prime+}\to)\pi^+\pi^0)&\approx& 2.60\times 10^{-4},\nonumber\\ 
{\rm interference \quad term} &\approx& -4.55\times 10^{-4}.
\end{eqnarray}
It is shown that,
compared with the other two individual branching ratios,
 the interference term is indeed large and negative,
which in turn leads to the clear dip in the region around $\omega\sim 1.6$ GeV as shown in Fig. ~\ref{brwa}.

\begin{table}[htbp!]
	\centering
	\caption{ The branching ratios (in units of $10^{-3}$) of the two body $B_c \to [\eta_c(1S),J/\psi]V $ decays,
with $V=\rho, K^*$, in which the sources of theoretical errors are same as in
Table \ref{brthree}.
For a comparison,
the previous PQCD predictions \cite{Zhang:2025yvt,Liu:2023kxr,Rui:2014tpa} and the results from other theoretical approaches \cite{Colangelo:1999zn,Nayak:2022qaq,Chang:1992pt,Kiselev:2000pp,Hernandez:2006gt,Issadykov:2018myx,Cheng:2021svx,Zhang:2023ypl,
Liu:1997hr,AbdEl-Hady:1999jux,Qiao:2012hp,Ke:2013yka,
Kar:2013fna,Naimuddin:2012dy,Ivanov:2006ni,Ebert:2003cn,
Wang:2007sxa,Chang:2014jca,S:2024adt} are also listed.}
\begin{ruledtabular}
\begin{threeparttable}
	\begin{tabular}{lccccccccccc}
Channels                                              &${\rm This \quad Work }$&\cite{Zhang:2025yvt,Liu:2023kxr}&\cite{Rui:2014tpa}&\cite{Qiao:2012hp}&\cite{Zhang:2023ypl}
&\cite{Issadykov:2018myx}
&\cite{Naimuddin:2012dy,Kar:2013fna}      &\cite{Ivanov:2006ni} &\cite{AbdEl-Hady:1999jux} &\cite{Kiselev:2000pp}&\cite{Ebert:2003cn}\\\hline
$B_c^+ \to \eta_c(1S) \rho^+ $              &$4.80^{+1.38 }_{-0.94 }$&$5.37^{+1.46 }_{-1.12 }$&$9.83^{+3.33}_{-2.59}$ &$14.5^{+2.00}_{-3.07}$ &$6.01^{+2.00}_{-3.07}$
&$5.18\pm 1.04$
&$1.06$ &$4.5$  &$3.3$&$4.2$&$2.1$\\
$B_c^+ \to \eta_c(1S) K^{*+} $              &$0.26^{+0.10 }_{-0.05  }$&$0.31^{+0.08 }_{-0.07  }$&$0.57^{+0.16}_{-0.12}$  &$0.77^{+0.11}_{-0.17}$ &$0.34^{+0.01}_{-0.01}$
&$0.29\pm 0.06$
&$0.06$ &$0.25$ &$0.18$&$0.20$&$0.11$\\
$B_c^+ \to J/\psi \rho^+ $              &$3.15^{+0.90 }_{-0.59 }$&$3.69^{+1.06 }_{-0.81 }$&$8.20^{+1.68}_{-1.53}$&$8.08^{+1.18}_{-1.41}$ &$5.34^{+0.74}_{-0.64}$
&$3.34\pm 0.67$&$1.8$ &$4.9$  &$3.1$ &$4.0$&$1.6$\\
$B_c^+ \to J/\psi K^{*+} $              &$0.18^{+0.05 }_{-0.03  }$&$0.22^{+0.07 }_{-0.05  }$ &$0.48^{+0.11}_{-0.09}$&$0.43^{+0.06}_{-0.08}$&$0.31^{+0.04}_{-0.04}$
&
$0.19\pm 0.04$&$0.09$&$0.28$ &$0.18$&$0.22$&$0.10$\\ \hline
Channels                                              &\cite{Liu:1997hr}  &\cite{Chang:1992pt}
&\cite{Hernandez:2006gt}
&\cite{Cheng:2021svx}
&\cite{Wang:2007sxa} &\cite{Chang:2014jca}&\cite{Nayak:2022qaq}&\cite{S:2024adt} &\cite{Colangelo:1999zn}&\cite{Ke:2013yka}\\\hline
$B_c^+ \to \eta_c(1S) \rho^+ $             &$3.51$&$5.9$
&$2.4$&$\cdots$
&$\cdots$&$\cdots$&$1.24$& $3.91^{+0.17}_{-0.27}$&$0.67$&$\cdots$\\
$B_c^+ \to \eta_c(1S) K^{*+} $              &$0.18$&$0.31$
&$0.13$&$\cdots$
&$\cdots$&$\cdots$&$0.065$& $0.19^{+0.01}_{-0.01}$&$\cdots$&$\cdots$\\
$B_c^+ \to J/\psi \rho^+ $               &$2.9$ &$6.5$
&$2.4$&$7.18^{+0.29}_{-0.33}$
&$5.0^{+2.0}_{-1.7}$&$3.20^{+0.58}_{-0.51}$&$\cdots$&$\cdots$&$3.7$&$0.109^{+0.033}_{-0.033}$\\
$B_c^+ \to J/\psi K^{*+} $              &$0.16$ &$0.35$
&$0.14$&$0.43\pm 0.01$ &$0.29^{+0.11}_{-0.10}$&$0.201^{+0.026}_{-0.025}$&$\cdots$&$\cdots$&$\cdots$&$\cdots$\\
\end{tabular}
\end{threeparttable}
\end{ruledtabular}
\label{brtwo1S}
\end{table}

\begin{table}[htbp!]
	\centering
	\caption{ The branching ratios (in units of $10^{-3}$) of the two body $B_c \to [\eta_c(2S),\psi(2S)]V $ decays,
with $V=\rho, K^*$, in which the sources of theoretical errors are same as in
Table \ref{brthree}.
For a comparison,
the previous predictions  from other theoretical approaches \cite{Zhang:2023ypl,Ebert:2003cn,Bediaga:2011cs,Qiao:2012hp,Liu:1997hr,Chang:1992pt,Nayak:2022qaq,Chang:2014jca,Colangelo:1999zn,Ke:2013yka} are also listed.
For the covariant light-front approach \cite{Zhang:2023ypl},
the results in both  scenario I   and scenario II (in the parentheses) are quoted. }
\begin{ruledtabular}
\begin{threeparttable}
	\begin{tabular}{lcccccccccccccc}
Channels                                              &${\rm This \quad Work }$&\cite{Zhang:2023ypl}&\cite{Ebert:2003cn}&\cite{Bediaga:2011cs}     &\cite{Qiao:2012hp}  &\cite{Liu:1997hr} \\\hline
$B_c^+ \to \eta_c(2S) \rho^+ $              &$0.84^{+0.26 }_{-0.18 }$
&$0.854^{+0.293 }_{-0.227 }(0.117^{+0.079 }_{-0.053 })$&$0.36$&$0.55$&$\cdots$ &$0.525$  \\
$B_c^+ \to \eta_c(2S) K^{*+} $              &$0.042^{+0.014 }_{-0.008  }$
&$0.049^{+0.017 }_{-0.013 }(0.0067^{+0.0045 }_{-0.0030 })$ &$0.02$&$0.026$&$\cdots$ &$0.025$ \\
$B_c^+ \to  \psi(2S) \rho^+ $              &$0.85^{+0.23 }_{-0.19 }$ &$1.193^{+0.405 }_{-0.315 }(0.718^{+0.284 }_{-0.215 })$ &$0.18$ &$1.1$&$2.11^{+0.30}_{-0.37}$ &$0.164$\\
$B_c^+ \to  \psi(2S) K^{*+}$              &$0.045\pm 0.011$
&$0.070^{+0.020 }_{-0.021 }(0.042^{+0.015 }_{-0.015 })$
&$0.01$
&$0.057$
&$0.112^{+0.016}_{-0.021}$&$0.0081$
 \\ \hline
Channels                                              &\cite{Chang:1992pt}&\cite{Nayak:2022qaq}&\cite{Chang:2014jca}    &\cite{Colangelo:1999zn}&\cite{Ke:2013yka}   \\\hline
$B_c^+ \to \eta_c(2S) \rho^+ $              &$0.668$&$\cdots$&$\cdots$&$0.15$&$\cdots$\\
$B_c^+ \to \eta_c(2S) K^{*+} $              &$0.033$&$\cdots$&$\cdots$&$\cdots$&$\cdots$\\
$B_c^+ \to  \psi(2S) \rho^+ $              &$0.765$&$1.246$ &$0.683^{+0.080}_{-0.110}$&$0.48$&$\cdots$\\
$B_c^+ \to  \psi(2S) K^{*+}$
&$0.046$
&$0.024$
&$0.041^{+0.003}_{-0.005}$
&$\cdots$&$0.044\pm 0.007$
\end{tabular}
\end{threeparttable}
\end{ruledtabular}
\label{brtwo2S}
\end{table}
\begin{table}[htbp!]
	\centering
	\caption{The branching ratios (in units of $10^{-3}$) of the two body $B_c \to \Psi V $ decays, with $V=\rho^{\prime},\rho^{\prime\prime}$,
in which the sources of theoretical errors are same as in
Table \ref{ebrthree}.
}
\begin{ruledtabular}
\begin{threeparttable}
	\begin{tabular}{lclc}
Channels                                              & PQCD predictions&Channels   & PQCD predictions               \\\hline
$B_c^+ \to J/\psi\rho^{\prime+}$              &$4.42^{+1.85}_{-1.45 }$ &$B_c^+ \to \eta_c(1S)\rho^{\prime+}$&$4.38^{+1.84 }_{-1.41 }$\\
$B_c^+ \to J/\psi\rho^{\prime\prime+}$              &$3.21^{+1.36 }_{-1.12 }$ &$B_c^+ \to \eta_c(1S)\rho^{\prime\prime+}$&$2.50^{+1.06}_{-0.84 }$\\
$B_c^+ \to \psi(2S)\rho^{\prime+}$              &$0.88^{+0.35 }_{-0.27 }$ &$B_c^+ \to \eta_c(2S)\rho^{\prime+}$&$0.52^{+0.22 }_{-0.18 }$ \\
$B_c^+ \to \psi(2S)\rho^{\prime\prime+}$              &$0.52^{+0.22 }_{-0.18 }$  &$B_c^+ \to \eta_c(2S)\rho^{\prime\prime+}$&$0.22^{+0.09}_{-0.06}$\\
\end{tabular}
\end{threeparttable}
\end{ruledtabular}
\label{ebrtwo}
\end{table}
The two-body branching ratio ${\cal B}(B_c^+ \to \Psi V)$ can usually be extracted from the corresponding quasi-two-body decay modes in Table \ref{brthree}
under the narrow width approximation:
\beq
\label{2body}
\mathcal{B}(B_c^+ \to \Psi (V\to)P_1P_2)&\approx& \mathcal{B}(B_c^+ \to \Psi V)\cdot
\mathcal{B}(V\to P_1P_2),
\eeq
with the assumption ${\cal B}(\rho^+\to \pi^+\pi^0)\approx 1$, ${\cal B}(K^{*+}\to K\pi)\approx 1$
and the estimates of ${\cal B}(\rho^{\prime+}\to\pi^+\pi^0)=(10.04^{+5.23}_{-2.61})\% $, ${\cal B}(\rho^{\prime\prime+}\to\pi^+\pi^0)=(8.11^{+2.22}_{-1.47})\%$ in Ref. \cite{Li:2017mao}.
So far,
we are not able to extract the two-body branching ratios ${\cal B}(B^+_c\to \Psi \rho^{\prime\prime\prime+})$ due to the lack information of ${\cal B}(\rho^{\prime\prime\prime+}\to \pi^+\pi^0)$.
For the strong decays of $I=1/2$ resonance $K^{*+}$ to $K\pi$,
the  isospin conservation is assumed
when we compute the ${\cal B}$ of the quasi-two-body process $B_c^+\to \Psi (K^{*+}\to)K^0\pi^+$,
\begin{eqnarray}
 \frac{{\cal B}(K^{*+}\to K^0\pi^+)}{{\cal B}(K^{*+}\to K\pi)}=\frac{2}{3},\quad
 \frac{{\cal B}(K^{*+}\to K^0\pi^0)}{{\cal B}(K^{*+}\to K\pi)}=\frac{1}{3}.
\end{eqnarray}
The obtained ${\cal B}$ of the two-body decays $B_c^+ \to \Psi V$ are then summarized in Tables \ref{brtwo1S}-\ref{ebrtwo},
in which the sources of theoretical errors are the same as in Tables \ref{brthree} and \ref{ebrthree}.
For a comparison,
the previous PQCD predictions \cite{Liu:2023kxr,Zhang:2025yvt,Rui:2014tpa} and the results from other theoretical approaches \cite{Colangelo:1999zn,Nayak:2022qaq,Chang:1992pt,Kiselev:2000pp,Hernandez:2006gt,Issadykov:2018myx,Cheng:2021svx,Zhang:2023ypl,
Liu:1997hr,AbdEl-Hady:1999jux,Qiao:2012hp,Ke:2013yka,S:2024adt,
Kar:2013fna,Naimuddin:2012dy,Ivanov:2006ni,Bediaga:2011cs,Ebert:2003cn,
Wang:2007sxa,Chang:2014jca} are also listed in Tables \ref{brtwo1S} and \ref{brtwo2S}.
For the four decays $B_c^+\to J/\psi \rho^+$,
$B_c^+\to J/\psi K^{*+}$ and $B_c^+\to \eta_c(1S) \rho^+$, $B_c^+\to \eta_c(1S)  K^{*+}$ in Table \ref{brtwo1S},
one can see that our calculations of the ${\cal B}$
are consistent well with the earlier PQCD predictions \cite{Liu:2023kxr,Zhang:2025yvt} within errors,
but a bit smaller than those given in Ref. \cite{Rui:2014tpa}.
The primary reason for this discrepancy is due to the choice of non-perturbative inputs,
such as the $B_c$ meson DA.
As shown in Eq.~(\ref{bcda}),
an improved version of $B_c$ meson LCDA has been adopted in our calculations,
where both the momentum fraction $x$ and impact parameter $b$ are taken into account.
In Ref. \cite{Liu:2023kxr},
the authors have taken the same form of LCDA as the present work.
However,
the authors in \cite{Rui:2014tpa} have taken
the zero-point $B_c$ LCDA
($\phi_{B_c}\propto \delta(x-r_c)$).
Although the individual ${\cal B}$ of the considered charmonium $B_c$ meson decays have not been measured so far,
the PQCD results of the
${\cal B}(B_c \to [\eta_c(1S), J/\psi] V) $ in Table \ref{brtwo1S} are generally comparable with most of the predictions from other theoretical  methods within errors,
in particular, with those given in
the covariant confined quark model \cite{Issadykov:2018myx}.
Furthermore, our calculations about the
$B_c\to [\eta_c(2S), \psi(2S)] V$ branching fractions in Table \ref{brtwo2S} are also close to those presented in  the Bethe
Salpeter equation \cite{Chang:1992pt} and   the covariant light-front approach (scenario I) \cite{Zhang:2023ypl}.
We hope that more precise data from the future LHCb experiments can help us differentiate these theoretical approaches and understand the
underlying mechanism of the charmonium $B_c$ meson hadronic weak decays.

We have also examined the effects on the branching ratios of the quasi-two-body $B_c\to \Psi(V\to)P_1P_2$ decays
from the introduced parameters $N_{K\pi}$ and $N_{\pi\pi}$ in Eqs.~(\ref{BRW}) and (\ref{GS}).
For example,
we recalculate the $\cal B$ of the two typical channels $B^+_c\to J/\psi(\rho^+\to)\pi^+\pi^0$ and
$B^+_c\to J/\psi(K^{*+}\to) K^0\pi^+$ with the fitted Gegenbauer coefficients in Ref.~\cite{Li:2020zng}:
${\cal B}(B^+_c\to J/\psi(\rho^+\to)\pi^+\pi^0)=2.91\times 10^{-3}$ and  ${\cal B}(B^+_c\to J/\psi(K^{*+}\to) K^0\pi^+)=0.06 \times 10^{-3}$,
in which the parameter $N_{K\pi(\pi\pi)}=1$ has been assumed.
According to the Eq.~(\ref{2body}),
the relevant two-body branching ratio ${\cal B}(B^+_c\to J/\psi\rho^+)$ and ${\cal B}(B^+_c\to J/\psi K^{*+})$
 are then estimated to be $2.91\times 10^{-3}$ and $0.09 \times 10^{-3}$, respectively.
Compared with the results in Table \ref{brtwo1S},
one can see that the ${\cal B}(B^+_c\to J/\psi\rho^+)$ is actually not sensitive to the parameter $N_{\pi\pi}$.
While for the $B^+_c\to J/\psi K^{*+}$ decay, on the other hand,
the obtained ${\cal B}=0.09 \times 10^{-3}$ becomes about twice smaller than the previous two-body PQCD calculation $0.22\times 10^{-3}$ \cite{Liu:2023kxr}.
As claimed previously,
the dominate contributions of the $B^+_c\to J/\psi\rho^+$ and $B^+_c\to J/\psi K^{*+}$
decays are from the longitudinal amplitude $F_{e\psi}^{LL,0}$ in Eq.~(\ref{f0epsi}),
which in fact has the relation of $F_{e\psi}^{LL,0}\propto F(\omega^2)\propto N_{K\pi(\pi\pi)}$.
The above phenomenon is then not hard to realize by comparing the two parameters $N_{\pi\pi}=1.05$ and $N_{K\pi}=1.48$,
in which the $N_{K\pi}=1.48$ shows a clear deviation from the unit.
Anyway,
the consistency between the quasi-two-body and two-body analyses of
the $B^+_c\to J/\psi (K^{*+}\to)K^0\pi^+$ and $B^+_c\to J/\psi K^{*+}$ decays, respectively, can be improved significantly in the PQCD approach,
when the effect from the parameter $N_{K\pi}$ has been taken into account in the calculation.
In addition,
the resultant Gegenbauer expansion of the $K\pi$ DAs
 becomes more convergent with the inclusion of the additional parameter $N_{K\pi}$ as shown in Eq.~(\ref{gegen}).

\begin{figure}[tbp]
\centerline{\epsfxsize=15cm \epsffile{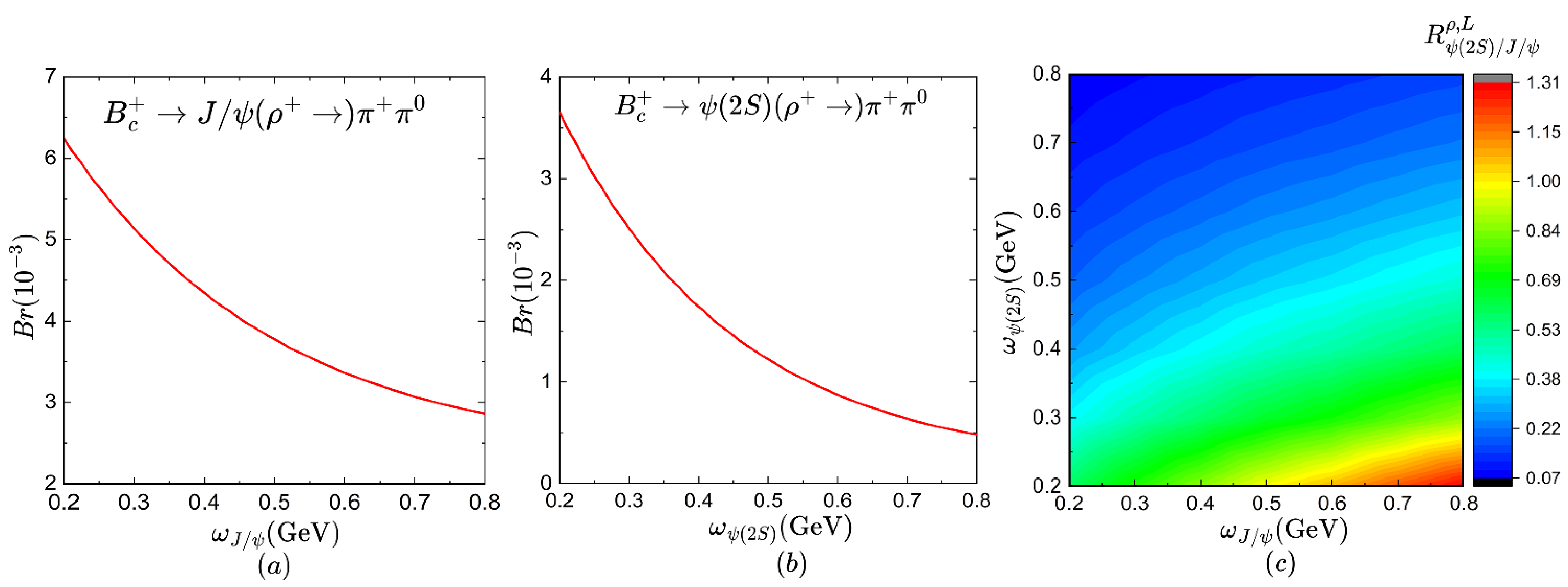}}
\vspace{-0.3cm}
\caption{Dependence of the (a) ${\cal B}(B^+_c\to J/\psi (\rho^+\to)\pi^+\pi^0)$, (b) ${\cal B}(B^+_c\to \psi(2S) (\rho^+\to)\pi^+\pi^0)$, and (c) relative ratio $R^{\rho, L}_{\psi(2S)/J/\psi}$ on the shape parameters in the charmonium DAs .}
\label{ww12}
\end{figure}
The decay mode $B_c^+\to \psi(2S)\pi^+$ has already been observed by the LHCb Collaboration with the measured ratio of the branching fractions as \cite{pdg2024}:
\begin{eqnarray}
R^{\pi, \rm exp}_{\psi(2S)/J/\psi}=\frac{{\cal B}(B_c^+\to \psi(2S)\pi^+)}{{\cal B}(B_c^+\to J/\psi\pi^+)}=0.254\pm 0.019.
\end{eqnarray}
Because the contributions from the factorizable emission diagrams dominate the $B_c^+\to J/\psi \pi^+ (\rho^+)$ and $B_c^+\to \psi(2S) \pi^+ (\rho^+)$ decays,
the following new defined ratio $R^{\rho, L}_{\psi(2S)/J/\psi}$
is expected to be close to the above measurement $R^{\pi, \rm exp}_{\psi(2S)/J/\psi}=0.254\pm 0.019$,
\begin{eqnarray}
R^{\rho, L}_{\psi(2S)/J/\psi}&=&\frac{{\cal B}(B_c^+\to \psi(2S)\rho^+) f_0(B_c^+\to \psi(2S)\rho^+)}{{\cal B}(B_c^+\to J/\psi\rho^+) f_0(B_c^+\to J/\psi \rho^+)}.
\end{eqnarray}
In Fig. \ref{ww12},
we have explored the influence on the ${\cal B}(B_c^+\to J/\psi( \rho^+\to)\pi^+\pi^0)$, ${\cal B}(B_c^+\to \psi(2S)( \rho^+\to)\pi^+\pi^0)$, and the relative ratio $R^{\rho, L}_{\psi(2S)/J/\psi}$ from the shape parameters in the charmonium DAs,
which manifest the sensitivity to $\omega_{J/\psi}$ and $\omega_{\psi(2S)}$.
In principle,
the non-perturbative parameters $\omega_{J/\psi}$ and $\omega_{\psi(2S)}$ in the $J/\psi$ and $\psi(2S)$ DAs
should also be constrained  from the global analysis of the multi-body charmonium $B_c$ meson decays.
Due to the limited experimental data,
however,
it is not practical for us to perform the global fitting at present.
For simplicity,
we shall adopt the latest extracted  $\omega_{J/\psi}=0.667\pm 0.080$ GeV for the $J/\psi $ meson~\cite{Dey:2025xdx}, which is obtained by fitting the PQCD form factors of $B_c \to J/\psi$ transition with corresponding lattice input.

\begin{figure}[tbp]
\centerline{\epsfxsize=9cm \epsffile{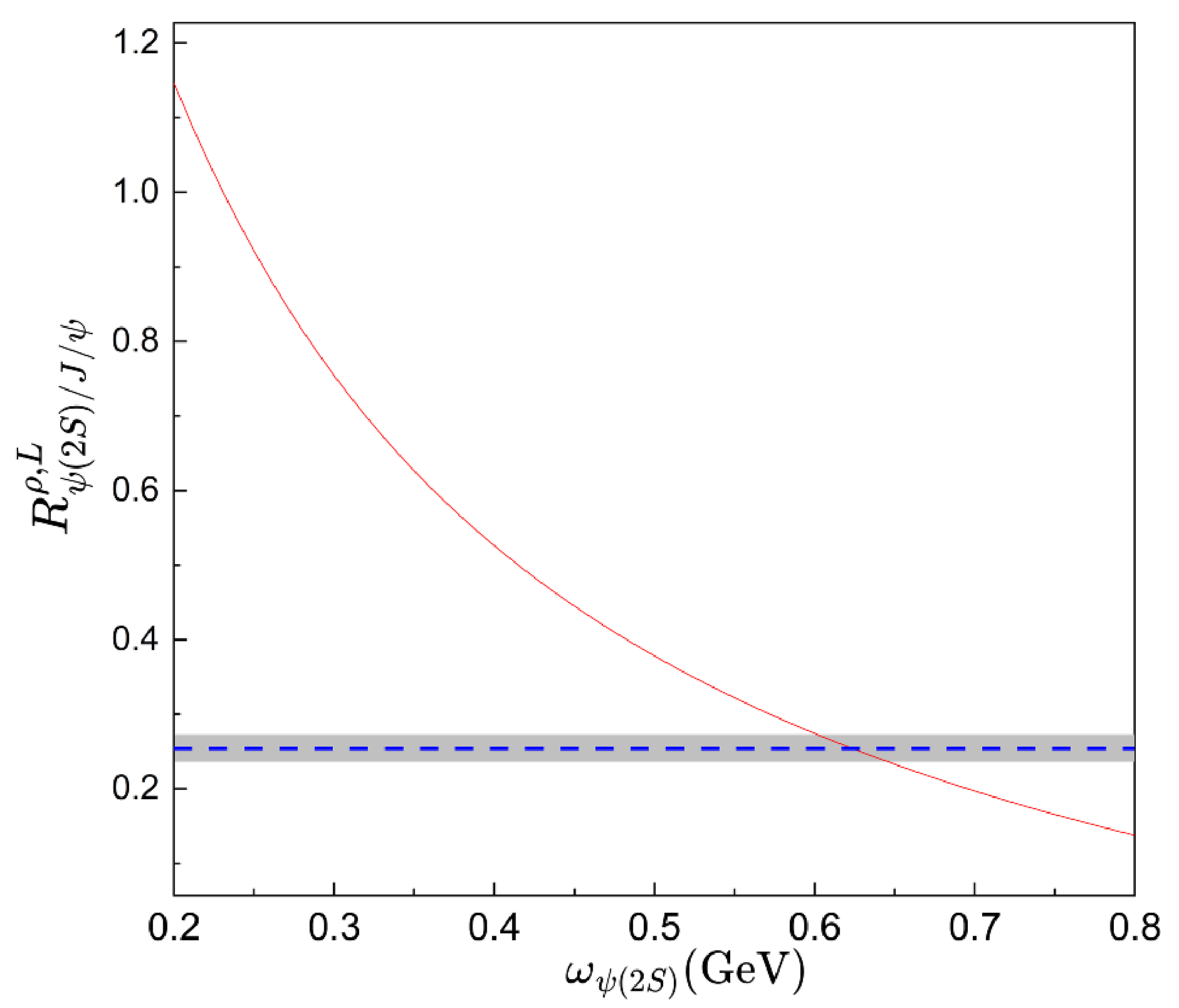}}
\vspace{-0.3cm}
\caption{Dependence of the ratio $R^{\rho,L}_{J/\psi/\psi(2S)}$ on the shape parameter $\omega_{\psi(2S)}$, where $\omega_{J/\psi}=0.667$ GeV \cite{Dey:2025xdx} is fixed. The
dashed horizontal line corresponds to the central value of the data \cite{pdg2024}
with the shaded band representing the experimental errors.}
\label{Rjpsi}
\end{figure}

For the radially excited state $\psi(2S)$ meson,
the value of $\omega_{\psi(2S)}$ can not be obtained in the similar proposal in \cite{Dey:2025xdx}
because the lattice calculation for the $B_c \to \psi(2S)$ transition form factors are not available yet.
The $\omega_{\psi(2S)}=0.20\pm 0.10$ GeV\cite{Rui:2015iia} is usually chosen in the previous PQCD analysis of the charmonium $B_c$ meson decays,
such that the valence charm quark, carrying the invariant mass
$x^2p^2\sim m_c^2$, is almost on shell.
Nonetheless,
it is found that the predicted ratio $R^{\rho, L}_{\psi(2S)/J/\psi}=1.15$ could be much larger than the experimental data $R^{\pi, \rm exp}_{\psi(2S)/J/\psi}=0.254\pm 0.019$ \cite{pdg2024} when   $\omega_{\psi(2S)}=0.20\pm 0.10$ GeV is used  in our calculation.
Since one of the peaks of the
$\psi(2S)$ DA is so close to the peak
of the $B_c$ meson DA, the overlaps between them are large when $\omega_{\psi(2S)}=0.20$ GeV, which can enhance the value of the calculated  ${\cal B}(B_c^+\to \psi(2S)( \rho^+\to)\pi^+\pi^0)$ and then lead to the large ratio  $R^{\rho, L}_{\psi(2S)/J/\psi}$ spontaneously.
Therefore,
we set $\omega_{\psi(2S)}$ as a free parameter in the current work, and plot the ratio $R^{\rho, L}_{\psi(2S)/J/\psi}$ dependent on it in Fig.~\ref{Rjpsi},
where $\omega_{J/\psi}=0.667$ GeV~\cite{Dey:2025xdx} is fixed.
Combined with the experimental data,
the value of the $\omega_{\psi(2S)}$ can then be bounded in the range of $\omega_{\psi(2S)}=0.62\pm 0.02$ GeV,
in which way the predicted $R^{\rho,L}_{\psi(2S)/J/\psi}=0.259^{+0.002}_{-0.016}$ becomes matching well with the data $0.254\pm 0.019$ \cite{pdg2024}.
In addition,
we also assume $\omega_{\eta_c(2S)}=\omega_{\psi(2S)}$ in our work because of the limited information on the $\eta_c(2S)$ meson.

In order to calculate the relative ratio $R_{2\pi/\pi}$ mentioned in the Introduction,
we have recalculated the ${\cal B}$ of the two-body decay $B_c^+\to J/\psi \pi^+$ with the same inputs as the current work,
\begin{eqnarray*}
{\cal B}(B_c^+\to J/\psi \pi^+)=(1.18^{+0.34}_{-0.26})\times 10^{-3},
\end{eqnarray*}
where the sources of theoretical errors are
from the variations of the shape parameter $\beta_{B_c}$ in the $B_c$ meson DA, of the shape parameter $\omega_{J/\psi}$ in the $J/\psi$ meson DA,
of the Gegenbauer moments in the $\pi$ meson DA and of the hard scale $t$,
but added in quadrature.
The ratio $R_{2\pi/\pi}$ can then be evaluated as follows,
\begin{eqnarray}
R^{\rm PQCD}_{2\pi/\pi}=\frac{{\cal B}(B_c^+\to J/\psi (\rho^+\to)\pi^+\pi^0)}{{\cal B}(B_c^+\to J/\psi\pi^+)}=2.67^{+0.26}_{-0.18},
\end{eqnarray}
which is consistent well with the latest LHCb measurement \cite{LHCb:2024nlg} within errors,
\begin{eqnarray}
R^{\rm exp}_{2\pi/\pi}=\frac{{\cal B}(B_c^+\to J/\psi \pi^+\pi^0)}{{\cal B}(B_c^+\to J/\psi\pi^+)}=2.80\pm 0.25.
\end{eqnarray}
Similarly,
one can also define the following ratios of ${\cal B}(B^+_c\to \psi(2S)(\rho^+\to)\pi^+\pi^0 )$, ${\cal B}(B^+_c\to J/\psi(\psi(2S)) (K^{*+}\to)K^+\pi^0 )$
${\cal B}(B^+_c\to \eta_c(1S, 2S) (\rho^+\to)\pi^+\pi^0 )$ and ${\cal B}(B^+_c\to \eta_c(1S, 2S) (K^{*+}\to)K^+\pi^0 )$ over
${\cal B}(B_c^+\to J/\psi\pi^+)$:
\begin{eqnarray}
R^{\psi(2S)}_{2\pi/\pi}&=&0.72^{+0.06}_{-0.06},\quad R^{\eta_c(1S)}_{2\pi/\pi}=4.07^{+0.42}_{-0.32},\quad R^{\eta_c(2S)}_{2\pi/\pi}=0.71^{+0.07}_{-0.05},\nonumber\\
R^{J/\psi}_{K\pi/\pi}&=&0.10^{+0.01}_{-0.01} , \quad R^{\psi(2S)}_{K\pi/\pi}=0.025^{+0.002}_{-0.006}, \quad R^{\eta_c(1S)}_{K\pi/\pi}=0.14^{+0.03}_{-0.01},
\quad R^{\eta_c(2S)}_{K\pi/\pi}=0.024^{+0.002}_{-0.001}.
\end{eqnarray}
The above seven new ratios are expected to be verified by the LHCb collaboration in the near future.

The relative ratios between the $CP$-averaged branching fractions of the
$B_c^+\to J/\psi K^{*+}$ and $B_c^+\to J/\psi \rho^+$ decays can be obtained as follows:
\begin{eqnarray}\label{ksrho}
R^{J/\psi}_{K^*/\rho}&=&\frac{{\cal B}(B_c^+\to J/\psi K^{*+})}{{\cal B}(B_c^+\to J/\psi \rho^+)}=0.057^{+0.005}_{-0.006}.
\end{eqnarray}
It is interesting to note that
the ratio defined in Eq. (\ref{ksrho}) from most of the theocratical methods \cite{Wang:2007sxa,Zhang:2023ypl,Liu:2023kxr,Qiao:2012hp,Chang:1992pt,Liu:1997hr,AbdEl-Hady:1999jux,
Ivanov:2006ni,Hernandez:2006gt,Naimuddin:2012dy,Cheng:2021svx,Rui:2014tpa,Issadykov:2018myx,Kiselev:2000pp,Ebert:2003cn,Kar:2013fna} are actually close to each other,
though there exit clear discrepancy among the calculated $\cal B$ of the individual $B_c^+\to J/\psi\rho^+ $ and $B_c^+ \to J/\psi K^{*+} $ decays.
As aforementioned,
the $B_c^+\to J/\psi\rho^+, J/\psi K^{*+} $ decays are all predominated by the factorizable emission diagrams,
while the  non-factorizable emission amplitudes are rather small.
It means that the theoretical calculations of $R^{J/\psi}_{K^*/\rho}$ from various approaches should be consistent with each other,
which is naively anticipated in factorization ansatz.
In fact, as described in the naive factorization approach,
the total decay amplitudes of these channels could be approximately written into the product of decay constant and the corresponding transition form factors.
The ratio $R^{J/\psi}_{K^*/\rho}$ can then be naively further simplified as
\begin{eqnarray}\label{niveksrho}
R^{J/\psi, {\rm naive}}_{K^*/\rho} =\frac{{\cal B}(B_c^+\to J/\psi K^{*+})}{{\cal B}(B_c^+ \to J/\psi \rho^+)}\approx
\frac{|V_{us}|^2}{|V_{ud}|^2}\cdot\frac{f_{K^*}^2}{f_{\rho}^2}=0.058,
\end{eqnarray}
with $|V_{us}|=0.2265$, $|V_{ud}|=0.9740$, $f_{K^*}=0.217$ MeV, and $f_{\rho}=0.209$ MeV \cite{Ali:2007ff}.
One can see easily that the above naive expectation $R^{J/\psi, {\rm naive}}_{K^*/\rho} =0.058$ is consistent well with the result in Eq.  (\ref{ksrho}).
The same argument also applies to the ratios between the ${\cal B}$ of other $B_c^+\to \Psi K^{*+}$ and $B_c^+\to \Psi \rho^+$ decays, with $\Psi=\eta_c(1S),\eta_c(2S),\psi(2S)$,
\begin{eqnarray}
R^{\psi(2S)}_{K^*/\rho}=0.053^{+0.002}_{-0.012}, \quad
R^{\eta_c(1S)}_{K^*/\rho}=0.054^{+0.004}_{-0.004}, \quad
R^{\eta_c(2S)}_{K^*/\rho}=0.050^{+0.002}_{-0.003}.
\end{eqnarray}
Since the similar ratio $R^{J/\psi}_{K/\pi}={\cal }(B_c^+\to J/\psi K^+ )/{\cal }(B_c^+\to J/\psi \pi^+ )=0.079\pm 0.008$ \cite{pdg2024}
has already been confirmed by the LHCb collaboration,
the ratios  $R^{\Psi}_{K^*/\rho}$, with $\Psi=J/\psi, \psi(2S), \eta_c(1S), \eta_c(2S)$, are expected to be verified soon.

\section{CONCLUSION}
We have systematically studied the quasi-two-body $B_c \to \Psi (V\to)P_1P_2$ decays by employing the leading order  PQCD framework,
with the charmonium meson $\Psi=\eta_c(1S), \eta_c(2S), J/\psi, \psi(2S)$ and the meson pair $P_1P_2=\pi\pi, K\pi$.
The invariant mass spectra of the final-state $\pi\pi$ and $K\pi$ pairs are dominated by the $P$-wave resonance $\rho(770)$ and $K^*(892)$, respectively.
The contributions from the excited $\rho^{\prime}(1450)$, $\rho^{\prime\prime}(1700)$ and $\rho^{\prime\prime\prime}(2254)$ resonances are also concerned
in the $\pi\pi$  invariant-mass spectrum.
The strong dynamics associated with the hadronization of the final state meson pairs is parametrized into the non-perturbative two-meson DAs,
which include both resonant and nonresonant contributions and have been established in quasi-two-body $B$ meson decays.

With the updated two-meson DAs determined from our previous works,
the branching ratios and polarization fractions of the considered charmonium quasi-two-body decays $B_c\to \Psi (V\to) P_1P_2$ have been examined.
It is worthwhile to note that the direct $CP$ violations of the $B_c\to \Psi (V\to) P_1P_2$ decays are
naturally expected to be zero since only tree operators work on these decays.
The longitudinal polarization fractions of the $B_c\to [J/\psi,\psi(2S)](\rho^+(770)\to)\pi^+\pi^0$ and
$B_c\to [J/\psi,\psi(2S)](K^{*+}\to)K^0\pi^+$ decays are found to be as large as $\sim 90\%$,
because the transverse amplitudes from the dominant factorizable emission diagrams  are usually power suppressed compared to the longitudinal ones.

As a by-product, we extracted the two-body $B_c \to \Psi V$ branching ratios from the predictions for the
corresponding quasi-two-body modes under the narrow width approximation.
The obtained results of the $B_c^+ \to [\eta_c(1S), J/\psi] \rho^+(770)$ and $B_c^+ \to [\eta_c(1S), J/\psi] K^{*+}$ decays
agree well with the previous calculations performed in the two-body PQCD framework.
The consistency between the quasi-two-body and two-body analyses supports the PQCD approach to exclusive charmonium $B_c$ meson decays.

Besides,
we defined a series of relative ratios for properly selected pairs of the considered $B_c$ decay modes.
In particular,
 the calculated ratio $R^{\rm PQCD}_{2\pi/\pi}=2.67^{+0.26}_{-0.18}$  is consistent well with the currently available LHCb measurement $R_{2\pi/\pi}^{\rm exp}=2.80\pm 0.25$.
The predictions in this work need to be further tested by the future experiments.

\begin{acknowledgments}
This work is supported by the National Natural Science Foundation of China under
Grant Nos.~12005103, 12075086.
\end{acknowledgments}


\end{document}